\documentclass[11pt]{article}

\usepackage[margin = 1in]{geometry}
\usepackage{amsmath}
\usepackage{amssymb}
\usepackage{graphicx}
\usepackage{float}
\usepackage{caption}
\usepackage{amsthm}
\usepackage{thm-restate}
\usepackage{xcolor}
\definecolor{green}{rgb}{0,0.5977,0}
\definecolor{orange}{HTML}{FF7F00}
\usepackage[linesnumbered,algoruled,boxed,lined,noend]{algorithm2e}
\usepackage{enumerate}
\usepackage{url}
\usepackage{soul}

\newcommand{\rr}{\mathbb R}

\newcommand{\poly}{\mathrm{poly}}

\newcommand{\inv}{^{-1}}
\newcommand{\abs}[1]{\left|{#1}\right|}

\newcommand{\suchthat}{\ | \ }

\newcommand{\tth}{^\text{th}}

\newcommand{\twocases}[4]{\begin{cases} #2 & #1 \\ #4 & #3 \end{cases}}

\newcommand{\txt}[1]{\text{#1}}

\newcommand{\ipncm}[3]{\begin{figure}[H]\begin{center}\includegraphics[scale = {#1}]{Media/#2.pdf}\caption{#3}\end{center}\end{figure}}

\makeatletter 
\g@addto@macro{\@algocf@init}{\SetKwInOut{Parameter}{Parameters}} 
\makeatother

\makeatletter
\newcommand{\nosemic}{\renewcommand{\@endalgocfline}{\relax}}% Drop semi-colon ;
\newcommand{\dosemic}{\renewcommand{\@endalgocfline}{\algocf@endline}}% Reinstate semi-colon ;
\let\oldnl\nl% Store \nl in \oldnl
\newcommand{\nonl}{\renewcommand{\nl}{\let\nl\oldnl}}% Remove line number for one line
\makeatother

\newtheoremstyle{skinny}      
{3pt} %Aboveskip 
{3pt} %Below skip
{\itshape} %Body font e.g.\mdseries,\bfseries,\scshape,\itshape
{} %Indent
{\bfseries} %Head font e.g.\bfseries,\scshape,\itshape
{.} %Punctuation afer theorem header
{ } %Space after theorem header
{} %Heading
\newtheoremstyle{skinnydefn}      
{3pt} %Aboveskip 
{3pt} %Below skip
{} %Body font e.g.\mdseries,\bfseries,\scshape,\itshape
{} %Indent
{\bfseries} %Head font e.g.\bfseries,\scshape,\itshape
{.} %Punctuation afer theorem header
{ } %Space after theorem header
{} %Heading
\theoremstyle{skinny}
\newtheorem{theorem}{Theorem}

\numberwithin{theorem}{section} 

\newtheorem{lemma}[theorem]{Lemma}

\theoremstyle{skinnydefn}
\newtheorem{definition}[theorem]{Definition}

\newcommand{\wH}{\overline{H^*}}

\usepackage{titlesec}
\titlespacing*{\section}{0pt}{3ex plus 1ex minus .2ex}{2ex plus .2ex}
\titlespacing*{\subsection}{0pt}{2ex plus 1ex minus .2ex}{1.5ex plus .2ex}
\titlespacing*{\subsubsection}{0pt}{0.3\baselineskip}{0.2\baselineskip}

\title{Sampling Tree-Weighted Partitions Without Sampling Trees}\author{Sarah Cannon\footnote{Claremont McKenna College, \texttt{scannon@cmc.edu}}, Topher Pankow\footnote{Yale University, \texttt{topher.pankow@yale.edu}}, Wesley Pegden\footnote{Carnegie Mellon University, \texttt{wes@math.cmu.edu}}, and Jamie Tucker-Foltz\footnote{Yale University, \texttt{j.tuckerfoltz@yale.edu}}}

\begin{document}

\maketitle
\thispagestyle{empty}

\begin{abstract}

This paper gives a new algorithm for sampling tree-weighted partitions of a large class of planar graphs. The tree-weighted distribution on $k$-partitions of a graph weights $k$-partitions proportional to the product of the number of spanning trees of each part. Equivalently, it is the distribution on $k$-partitions induced by the components of uniformly random $k$-component forests. Recent work on problems in computational redistricting analysis has driven special interest in the conditional distribution where all partition classes have the same size ({\it balanced} partitions). One class of Markov chains in wide use aims to sample balanced tree-weighted $k$-partitions using a sampler for balanced tree-weighted 2-partitions. Previous implementations of this 2-partition sampler would draw a random spanning tree and check whether it contains an edge whose removal produces a balanced 2-component forest; if it does, this 2-partition is accepted, otherwise the algorithm rejects and repeats. In practice, this step is a significant computational bottleneck.

We show it's possible to sample from this balanced tree-weighted 2-partition distribution directly. Our new algorithm can quickly accept or reject a potential tree-weighted 2-partition, without first sampling a spanning tree; acceptance and rejection rates are the same as previous samplers.  We prove that on a wide class of planar graphs (encompassing network structures typically arising from the geographic data used in computational redistricting), our algorithm takes expected linear time $O(n)$. This is asymptotically faster than the best known method to generate random spanning trees in graphs where $m = O(n)$, which is $O(n \log^2 n)$ for approximate sampling and $O(n^{1 + \log \log \log n / \log \log n})$ for exact sampling. Of independent interest, we also show  a variant of our algorithm  gives a speedup to $O(n \log n)$ for exact sampling of uniformly random trees on these families of graphs. This improved the bounds for both exact and approximate sampling. Our algorithms and proofs involve planar duality, geometric analysis of random walks, and careful choices of planar separators.

We implement our algorithm and benchmark it on grid graphs, finding that it outperforms the standard bipartitioning method in the widely-used \emph{GerryChain} library.

\end{abstract}
 
\newpage
\section{Introduction}

\setcounter{page}{1}

Across the world, regions are divided into districts for the purpose of electing representatives.  The ways these district lines are drawn can have enormous effects on who is elected. In computational redistricting, one goal is to randomly sample possible political districting plans. This is useful for understanding the space of possible plans and what is typical in certain settings, and can also be used to flag outliers as potential gerrymanders~\cite{chen-rodden-thicket,ReCom,DukeNC}.  Such methods have been used in numerous legal cases, including in briefs submitted to the U.S. Supreme Court~\cite{ chen2022brief, duchin2019brief, hirsch2022brief}.

The geography of a state can be represented as a graph whose nodes are small geographic units (e.g. census blocks or voting precincts), weighted by population, with edges representing geographic adjacency.  In this setting, a political districting plan with $k$ districts is a partition of this graph into $k$ connected pieces. Typically, districts must have similar population.  This corresponds to each piece of the partition having similar total weight, and partitions satisfying this are called {\it balanced}.  Additionally, districts are often required or preferred to be {\it compact}. One way this has been operationalized is to sample from a measure like the spanning tree distribution, which gives each $k$-partition weight proportional to the product of the number of spanning trees of each partition class~\cite{ForestReCom,  RevReCom, conjecturepaper, ReCom, smc}.  Formally, for a partition $P$ with classes $P_1,\dots,P_k$, this distribution is\vspace{-4pt}
  \[
  \pi(P)\propto \prod_{i=1}^k \tau(G[P_i]),\vspace{-4pt}
  \]
where $\tau(H)$ denotes the number of spanning trees of graph $H$ and $G[X]$ is the subgraph of $G$ induced by vertex set $X$.  Note this distribution is only supported on partitions where every partition class induces a connected subgraph, so that all $\tau(G[P_i])$'s are nonzero.  

 \vspace{7pt}   
{\bf Sampling Methods for Computational Redistricting.} For redistricting analyses, interest in the spanning tree distribution or closely related distributions has been driven both by geometric properties of its typical samples \cite{RevReCom, ReCom} and availability of promising routes to efficiently generate good samples.  Recent work showed polynomial time sampling is possible on grid-like graphs for fixed $k$, but with a strong dependence on~$k$~\cite{polytimeforests}; in practice, samples are still typically generated using Markov chains \cite{ForestReCom, forestrecommultiscale, VRA-ELJ, RevReCom,  chenstephanopoulos, va-criteria, ReCom, MRL} which lack rigorous mixing time bounds but frequently do well on heuristic mixing tests.

These Markov chains reduce the problem of generating a balanced tree-weighted $k$-partition to that of generating a balanced tree-weighted 2-partition.  The basic idea is to repeatedly choose pairs of partition classes whose union is connected, view these two classes as a balanced tree-weighted 2-partition of their union, and resample that 2-partition according to either the spanning tree distribution~\cite{RevReCom} or another closely related distribution, see below~\cite{ReCom}. Thus the computational bottleneck for carrying out one such transition is solving the balanced tree-weighted 2-partition problem.

Randomly sampling a balanced tree-weighted 2-partition is typically done by generating a uniformly random spanning tree and checking whether it can be split into two equal-sized pieces; if it can't, reject. This produces samples drawn from the following distribution: for a 2-partition with classes $P_1$ and $P_2$, if there are $|E(P_1, P_2)|$ edges in $G$ between $P_1$ and $P_2$, 
\[
\overline{\pi}(P) \propto \tau(G[P_1]) \cdot \tau(G[P_2]) \cdot |E(P_1, P_2)|
\]
This alternate distribution $\overline{\pi}$ is widely targeted in this resampling step~\cite{ReCom}, and is the focus of our work. If desired, samples can also be drawn exactly from $\pi$ by incorporating a second rejection step: If the new proposed partition has parts $P_1'$ and $P_2'$, accept this new split with probability $1/|E(P_1', P_2')|$, otherwise reject~\cite{RevReCom}.

\subsection{Our Contributions}\label{subContributions}

Algorithms to generate uniformly random spanning trees are slow enough to hamper implementation on large graphs. We provide a new approach for this key subroutine which does not require drawing an entire spanning tree. Instead, we are able to directly sample from $\overline{\pi}$ faster than the fastest known algorithm to generate a uniformly random spanning tree. 

Our main result is a linear time algorithm for sampling balanced tree-weighted 2-partitions on well-behaved planar graphs. We also give a linear time algorithm for sampling {\it approximately} balanced 2-partitions. Of independent interest, our approach additionally gives an improved running time for sampling complete spanning trees on these graphs. We begin by (informally) defining the set of graphs our results apply to. 

\vspace{4pt}
\noindent {\bf Grid-Like Graphs.} 
Formalized in Definition~\ref{dfnGridLikeLattice}, a {\it grid-like lattice} is an infinite plane graph~$\Lambda$ with dual plane graph $\Lambda^*$ satisfying: (1) Bounded degree in $\Lambda$ and $\Lambda^*$; (2) No long edges in $\Lambda^*$; (3)  Approximately uniform density of vertices in both $\Lambda$ and $\Lambda^*$; (4) Random walks in $\Lambda^*$ can go in any direction; (5) Random walks in $\Lambda^*$ escape sets quickly. 
All of our results apply to {\it grid-like graphs}, which are finite, simply-connected\footnote{The simple connectivity assumption is a potential limitation of the practical applicability of our runtime analysis. We discuss this in detail in Sec.~\ref{subNotsc}, where we include empirical evidence the assumption is not too strong.} subgraphs of a grid-like lattice. One canonical example of a grid-like lattice is $\mathbb{Z}^2$, but our results apply to a much larger class of graphs as well. 

This class of graphs is interesting and relevant because the graphs encountered in redistricting applications are often planar. Additionally, random graph models approximating such dual graphs rely on the presence of many short edges and nearest-neighbor connections~\cite{dgs}. This makes grids, and graphs with similar properties, the logical graph classes to consider in this setting. 

\vspace{4pt}
\noindent {\bf Exactly Balanced 2-Partitions.}
For a tree $T$, let $b(T)$ denote the (unique) balanced 2-partition obtained from $T$ by deleting an edge if such a partition exists, or $\bot$ otherwise.  If such a balance edge exists, it is unique, so there is at most one possible partition. Our main result is as follows. 

\begin{theorem} \label{thmExact2balanced} For an $n$-vertex subgraph $H$ of a grid-like graph, there is a linear time (in $n$) algorithm to sample $b(\tau)$, where $\tau$ denotes a uniformly random spanning tree of $H$. 
\end{theorem}
\noindent Note if a partition is produced by this sampler, it is drawn exactly from $\overline{\pi}$.  Our algorithm is presented as Algorithm~\ref{algMain}; a subroutine for picking parameters for each stage of the algorithm is Algorithm~\ref{algQ}.  We prove our algorithm exactly samples $b(\tau)$ in Theorem~\ref{thmCorrect}, and prove that it runs in linear time in Theorem~\ref{thmMainRuntimeBound}. That this is possible in linear time is surprising, since all known algorithms for generating an (even approximately) random spanning tree take time $\Omega(n \log n)$. We are thus able to generate the partition induced by splitting a random spanning tree without generating the tree itself. 

The probability with which $b(\tau)$ returns a partition and not $\bot$ remains unknown. In~\cite{polytimeforests}, this probability was bounded below by $\Omega(n^{-2})$ for $n$-vertex grid-graphs. Experiments suggest this probability is in fact closer to $\Omega(n^{-1})$ in the real-world graphs used in computational redistricting~\cite{dgs}. If rejection sampling is used to find a balanced partition, this would imply an expected $O(n)$ samples are needed to accomplish this. 
Our work does not comment on the probability with which $b(\tau)$ returns $\bot$ (i.e. rejects), but it does speed up the determination of whether this is the case.

\vspace{4pt}
\noindent {\bf Approximately Balanced 2-Partitions.}
For a graph with $n$ vertices, we say a graph partition is $q$-balanced if the number of vertices in each component is within $\pm q$ of the target size of $\frac{n}{2}$.
In contrast to the case of exact balance, there may be multiple edges whose removal gives a $q$-balanced partition. In this case, if one wants to sample from the spanning tree distribution, subsequent rejection steps are more complicated but can be accomplished if one knows the number of balanced splits and can sample one uniformly~\cite{RevReCom}. The following theorem says we can still obtain a linear-time algorithm as long as the imbalance is not too large.

\begin{restatable}{theorem}{thmApprox}\label{thmApprox}
    For any function $q(n) = O(n / \log n)$ and any grid-like graph with $n$ vertices, there is an $O(n)$-time algorithm that counts the number of $q$-balanced partitions that can be obtained from a random spanning tree, and samples one of them uniformly if any exist.
\end{restatable}

\vspace{4pt}
\noindent {\bf Spanning Trees Sampling.}
 Our approach for the above results, discussed more thoroughly in Sec.~\ref{secProofOverview}, is Divide-and-Conquer. We carefully choose starting points of Wilson's algorithm for drawing random spanning trees so that the graph gets divided into smaller regions.  We then show that an edge inducing a balanced split can only be found in one of those regions, so we do not have to recurse on all of them. We also show that with constant probability, after a linear number of steps, the region size decreases by a constant factor. 
 This yields an overall linear runtime to generate the partition induced by the random tree.
On the other hand, if we care about generating the entire tree itself, rather than just the partition it induces, we can recurse on all regions, giving the following result. 
\begin{theorem}\label{thmST}
  For all grid-like graphs with $n$ vertices, it is possible to exactly sample a uniformly random spanning tree in time $O(n \log n)$. 
\end{theorem}
\noindent This is faster than the best algorithms for sparse graphs, namely, $O(n \log^2 n)$ for approximately uniform sampling~\cite{anari2021forestsampling} and $O(n^{1 + \log \log \log n / \log \log n})$ for exact uniform spanning tree sampling \cite{Schild}. While it has been observed that sampling a random spanning tree in time $O(n \log n)$ is possible on grid graphs~\cite{polytimeforests}, this theorem extends that result to the broader class of all grid-like graphs. 

\vspace{4pt}
\noindent {\bf Emprical Validation.} In Sec.~\ref{secExperiments}, we implement our algorithm and benchmark it against the open-source \emph{GerryChain}~\cite{gerrychain} library, widely used in redistricting applications. While our main result ``only'' shaves a log factor of runtime, this translates into real efficiency improvements in practice. For the task of bipartitioning square grid graphs, our method scales better than the standard method using Wilson's algorithm, with dramatic improvements for grids of skinnier aspect ratio.

\vspace{4pt}
\noindent {\bf Algorithm Simplicity.}
Finally, our algorithm has a qualitative feature making it especially appealing for the task of redistricting: It is explainable in very simple terms. This is desirable for justifying an algorithm's use in, say, a court of law. One key motivation for targeting the spanning tree distribution is that we can efficiently sample from it and it admits a closed form description. Yet the existing algorithms for this task are themselves opaque, with no simple description that doesn't delve into graph-theoretic terminology. Our algorithm has the following two-sentence description:

\vspace{-5pt}
\begin{quote}
    \emph{Repeatedly put your pen down on the map and start randomly moving around until the region is divided into at least two large pieces. Then glue pieces together into two groups so that the populations are balanced, subdividing larger pieces with further random pen strokes as necessary.}
\end{quote}

\vspace{-5pt}
\noindent This is a conceptual and visual embodiment of ``blind'' redistricting. It is striking that such a seemingly unpredictable process can be designed to sample from a well-known benchmark distribution.

\subsection{Related work}\label{subLitReview}

\noindent {\bf Minimum Spanning Trees vs. Uniform Spanning Trees.}
In practice, it is faster to sample a random minimum spanning tree (MST) than a random uniform spanning tree.  This is done by assigning random weights to each edge and then using a standard MST algorithm (such as Kruskal) to obtain a tree. 
There are several drawbacks to using random MSTs: the resulting distribution over trees is not uniform~\cite{ModelsOfRandomSpanningTrees, TappMST, RevReCom};  there is empirical and theoretical evidence suggesting random MSTs are less likely to be splittable into population-balanced subtrees~\cite{ModelsOfRandomSpanningTrees}; and there is no analog of Kirchoff's matrix-tree theorem to allow Metropolis-Hastings type approaches or the implementation of reversibility.
Therefore, it is advisable to use uniformly random spanning trees instead.

\vspace{4pt}
\noindent {\bf Sampling Uniform Spanning Trees.}
In~\cite{Schild}, the author achieves fast algorithms for random spanning tree generation by adding `shortcutters' to the Aldous-Broder algorithm~\cite{aldous, broder}. On graphs with $m$ edges and $n$ vertices, this results in a runtime of $m^{1 + O(\log \log \log n / \log \log n)}$ for exact uniform tree sampling. For sparse planar graphs, where $m = O(n)$, this can be simplified to $O(n^{1 + \log \log \log n / \log \log n})$. The best known method for approximately sampling random spanning trees uses an up-down Markov chain that iteratively adds a random edge to a spanning tree and then removes a random edge of the resulting cycle~\cite{anari2021forestsampling}. This chain has a known mixing time of $O(n \log n)$, and using clever data structures each step can be implemented in amortized time $O(\log n)$, resulting in an overall runtime of $O(n \log^2 n)$. 

For dense graphs like $K_n$, there is an information-theoretic lower bound $\Omega(n \log n)$ on the runtime of sampling a random spanning tree. This is because the number of possible spanning trees means writing one down requires at least this many bits. However, our focus is on the opposite end of the spectrum, looking at planar grid-like graphs where $m = O(n)$. No non-trivial lower bound is known for sparse~graphs.

\vspace{4pt}
\noindent {\bf Wilson's Algorithm and Planar Duality.}
No linear time algorithm for exactly or approximately sampling spanning trees in grids or grid-like graphs is known.  For grid graphs, the best known approach uses Wilson's algorithm~\cite{wilson}. Wilson's algorithm draws a random spanning tree by designating an arbitrary vertex $r$ as the root, and initially $T = \{r\}$. While $T$ is not yet a complete spanning tree, a loop-erased simple random walk starting from an arbitrary vertex $v \notin T$ proceeds until it reaches a vertex of $T$. When it reaches $T$, this entire path is added to $T$. This produces a uniformly random spanning tree, and the running time is closely related to the mean commute time between $r$ and a random vertex $v \neq r$. 

As there is a bijection between primal and dual spanning trees in planar graphs (see Sec.~\ref{sec:dual}), one can choose to run Wilson's algorithm in either. For grids, it is most efficient to run Wilson's algorithm in the dual graph with root $r$ chosen to be the dual vertex corresponding to the outer face of the grid graph. In this case,  Wilson's algorithm runs time expected time $O(n \log n)$ for $n$-vertex grid graphs. This is easily proved using the characterization of the commute time in terms of effective resistance (see also~\cite{polytimeforests}). 

Theorem~\ref{thmST} extends this runtime bound to grid-like graphs, a much larger class. This is an improvement on the primal methods typically used for spanning tree sampling in practice~\cite{ForestReCom,RevReCom,ReCom,smc}, where the only known bound on the time to generate one random spanning tree is $O(n^2)$ (from a general bound on the runtime of Wilson's algorithm for planar graphs).  Our work clearly shows the benefits of sampling structures in the planar dual graph rather than the primal graph.

\vspace{4pt}
\noindent {\bf Sampling Balanced Tree-Weighted $k$-Partitions.}
Moving beyond sampling spanning trees, a natural extension is the problem of sampling a partition of a graph into $k$ balanced parts, according to the spanning tree distribution. 

Approximately sampling (not necessarily balanced) tree-weighted $k$-partitions is possible in time $O(n \log^2 n)$, again using the up-down Markov chain~\cite{anari2021forestsampling}. Charikar, Liu, Liu and Vuong~\cite{conjecturepaper} conjectured that for grid graphs on $n$ vertices, for $k=O(1)$,  a $1/\poly(n)$ fraction of $k$-component forests have balanced component sizes. 
This was confirmed in \cite{polytimeforests}, where it was shown to hold for a wide class of grid-like graphs.
In particular, this means on such grid-like graphs one can approximately sample balanced tree-weighted $k$-partitions in polynomial time, for $k=O(1)$, via the up-down chain and rejection sampling.
The authors of~\cite{polytimeforests} also give a polynomial time exact sampling algorithm for this problem, again when $k = O(1)$. This method uses Wilson's algorithm to draw a random spanning tree, checks if the spanning tree can be cut into balanced pieces, and if so applies an additional probability filter before accepting.  

Currently, these are the only known approaches to sample balanced tree-weighted $k$-partitions in polynomial time. However, there appears to be an exponential dependence on $k$~\cite{polytimeforests, dgs}. This means it's not practical for generating plans with many districts. Instead, various Markov chains are used~\cite{ForestReCom, forestrecommultiscale, VRA-ELJ, RevReCom,  chenstephanopoulos, va-criteria, ReCom, MRL}. This is despite an absence of proofs of irreducibility (that all possible plans are reachable with recombination moves) and upper bounds on the mixing times (the time until samples are drawn from a distribution close to the chain's stationary distribution).

Our main result is an improvement on a crucial subroutine used by the Recombination Markov chain (ReCom) ~\cite{ReCom} and its reversible variant RevReCom~\cite{RevReCom}. This subroutine merges two districts and resamples a new partition of their union from $\overline{\pi}$. 
The primary computational bottleneck
is that the obvious way to do this resampling involves generating a random spanning tree. This takes time $O(n^2)$ when primal sampling methods are used, as they typically are in practice, or time $O(n \log n)$ when dual sampling methods are used (as we prove for grid-like graphs).
In this paper, we adapt tools from \cite{polytimeforests} to show that as long as the union of the 2 districts is simply connected, a 2-partition can be sampled from $\overline{\pi}$ in linear time---in particular, we show how to generate a random tree-weighted 2-partition faster than we know how to sample a random spanning tree.

\vspace{4pt}
\noindent {\bf Potential speed-ups for other sampling methods.}
In the related Forest ReCom chain~\cite{ForestReCom}, each step necessitates sampling entire trees rather than sampling partitions.
Because of this, Theorem~\ref{thmExact2balanced} doesn't provide any speed-up for this chain.  However,
our approach improves the time needed to sample a single spanning tree -- which is still a subroutine of Forest ReCom -- from $O(n^2)$ to $O(n \log n)$ when dual spanning trees are instead sampled using our methods (Theorem~\ref{thmST}).

Sequential Monte Carlo~\cite{smc} is another method for sampling tree-weighted partitions.
However, a key subroutine splits a random spanning tree into pieces of certain sizes before forgetting the tree and keeping only the partition. Our approaches may lead to speed-ups here as well, though additional work is needed to handle splitting trees into not-{(approximately-)}equally-sized pieces.

\section{Preliminaries}\label{secPreliminaries}

All graphs we consider are undirected, but may have self-loops or multiedges unless otherwise~stated.

\subsection{Planar Duals}\label{sec:dual}

Our results focus on planar graphs. For a finite, connected, planar graph $G$, fix an embedding of $G$ in the plane with no edges crossing. The \emph{dual graph} of $G$ (with respect to the embedding) is the graph $G^*$ whose vertices are faces of $G$, with an edge between faces $a^*, b^* \in V(G^*)$ whenever they share a common boundary edge in $G$. Note we count the outer face of $G$ as a vertex of $G^*$.

\begin{figure}[h]\begin{center}\includegraphics[scale = 0.8]{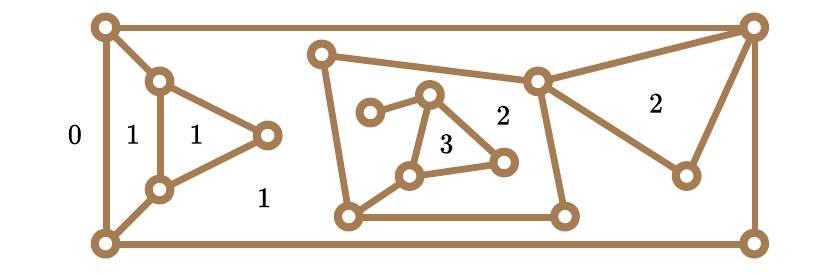}\caption{\label{figDepth}An embedded planar graph with the depth of each face labeled.}\end{center}\end{figure}

We define the \emph{depth} of each face $f \in V(G^*)$ to be the number of other faces entirely surrounding~$f$. Thus the outer face has depth 0, and all other faces have depth at least 1. Faces of depth 2 and greater can arise when there is some vertex $v$ whose removal would disconnect the graph into multiple components where one of these components is bounded by a cycle involving $v$, as in Fig.~\ref{figDepth}.

For a set of vertices $S \subseteq V(G)$, we say  the {\it boundary faces} of $S$ are all faces of $G$ that have some but not all of their incident vertices in $S$; these boundary faces correspond to vertices in $G^*$. For any edge $e \in E(G)$, let $e^* \in E(G^*)$ be the edge between the faces it bounds. For any set of edges $S \subseteq E(G)$, we define $S^* := \{e^* \suchthat e \in S\} \subseteq E(G^*)$. The following lemma is a standard result about planar graphs and their duals.

\begin{lemma}\label{lemDuality}
	Let $G$ be a connected planar graph, and fix a planar embedding. Then $e \mapsto e^*$ is a bijection between edges of $G$ and edges of $G^*$, and $T \mapsto T^* := (V(G^*), E(G^*) \setminus E(T)^*)$ is a bijection between spanning trees of $G$ and spanning trees of $G^*$.
\end{lemma}

\noindent Note that $T^*$ does not contain the edges $e^*$ for each $e \in T$, but rather those edges  {\it not} in this set.

\subsection{Grid-Like Lattices and Random Walks}\label{subLatticeSequences}

We now formally define the family of planar graphs to which our results apply.
The reader may find it easier to picture the square lattice throughout the remainder of the paper.

\begin{definition}\label{dfnGridLikeLattice}
A \emph{grid-like lattice} is an infinite simple plane graph $\Lambda$ (with a choice of embedding for the dual graph $\Lambda^*$) such that there exist positive constants $b, \rho, R, C, L$ so that:
\begin{enumerate}[(1)] \setlength{\itemsep}{0pt} 
\item\label{itmStrongLatticeSequenceDegree} \emph{Bounded degrees.} The degree of any vertex in $\Lambda$ or $\Lambda^*$ is at most $b$. 
    \item\label{itmStrongLatticeSequenceClose} \emph{No long dual edges.} Any adjacent pair $x, y\in V(\Lambda^*)$ satisfies $d(x, y) \leq 1$.
    \item\label{itmStrongLatticeSequenceDense} \emph{Approximately uniform density.} For any $p \in \rr^2$ and $r \geq R$, the ball $B_{r}(p)$ contains at most $C r^{2}$ vertices of $\Lambda$ and also at most $C r^{2}$ vertices of $\Lambda^*$.
    \item\label{itmStrongcircleproperty} \emph{Random dual walks can go any direction.} For any dual vertex $v \in \Lambda^*$ and any $r \geq R$, there is a division of the circle $C_{r,v}$ of radius $r$ centered around $v$ into arcs $A_1,\dots,A_s$, each of length at most $\frac 1 8 2\pi r$, such that the following holds for each $i \in [s]$: In a simple random walk $v = v_0, v_1, v_2, \dots$ in $\Lambda^*$, with probability at least $\rho > 0$, letting $j$ be the first index where $d(v_0, v_{j}) > r$, the straight line segment joining $v_{j - 1}$ with $v_j$ passes through $C_{r, v}$ in $A_i$.
    
    \item\label{itmStrongLatticeSequenceEscape} \emph{Random dual walks escape quickly.} For any set of dual vertices $S \subseteq \Lambda^*$, any $v \in S$, and any $p > 0$, the probability a random walk from $v$ takes more than $\frac{L\cdot\abs{S}}{p}$ steps to escape $S$ is at most~$p$.
\end{enumerate}
    
\end{definition}

\noindent One motivation for the reasonableness of (\ref{itmStrongLatticeSequenceDegree})--(\ref{itmStrongLatticeSequenceEscape}) is that the integer lattice satisfies them:
\begin{restatable}{proposition}{propzisagrid}\label{prop:zisagrid}
$\mathbb{Z}^2$ is a grid-like lattice.    
\end{restatable}
\noindent We give a proof in Appendix \ref{appz2isagrid}; the proof is easily adapted to other familiar lattices.

This definition intentionally mirrors the definition of ``uniform lattice sequences'' \cite{polytimeforests}, allowing us to import a key lemma analyzing the geometry of random walks on such graphs. Translated to our setting, the result is the following.

\begin{restatable}{lemma}{lemcurveswalk}\label{lem:curveswalk}
Let $\Lambda$ be any grid-like lattice, and let $R$ and $\rho$ be the constants from the definition of a grid-like lattice. For all sufficiently large $r$ (specifically, for $r \geq \max\{100, 10R\}$) and any curve $\gamma$ in the plane of positive length $|\gamma|$, with probability at least $\rho^{100\abs{\gamma}/r + 301}$ a random walk in $\Lambda^*$ from any $v_0\in \Lambda^*$ with $d(v_0,\gamma(0)) \leq r/2$ will eventually encircle $\gamma(1)$ while remaining distance at least $r/5$ from this point and never reach a vertex at distance $>r$ from $\gamma$.
\end{restatable}
\noindent A complete proof showing how this lemma follows from \cite[Lemma 4.4]{polytimeforests} is included in Appendix~\ref{appDeferredProofs}; here we give a proof sketch illustrating why the result is true. 

\vspace{3pt}
\noindent {\it Proof Sketch.}  At a high level, the result holds by considering a sequence of circles along $\gamma$. For intuition, we first explain an easier result: that the probability the walk goes from a point $v$ within distance $r/2$ of $\gamma(0)$ to some point within distance $r/2$ of $\gamma(1)$ without ever going distance farther than $r$ from $\gamma$ is at least  $\rho^{|\gamma|/20 + 1}$. Begin by considering the ball of radius $r/2$ centered at $v$, $B(v, r/2)$.  Curve $\gamma$ must exit this ball somewhere; look at which of the arcs $A_1, \ldots , A_s$  that $\gamma$ intersects.  With probability at least $\rho$, a random walk from $v$ first exits $B(v, r/2)$ across the same arc as $\gamma$ by~\ref{itmStrongcircleproperty}.  Because each arc has length at most $\frac18 2 \pi (r/2)$ and adjacent points along the walk are within distance 1 of each other by~\ref{itmStrongLatticeSequenceClose}, provided the ball is large enough ($r/2 > 10$) the first vertex of the walk outside $B(v, r/2)$ must be within distance $r/2$ of $\gamma$. Because the walk never left $B(v, r/2)$ prior to this, it never went more than distance $r$ from $\gamma$. One can prove progress of at least distance $r/20$ along $\gamma$ in each step except possibly the first, meaning $20|\gamma|/r + 1$ iterations are enough to traverse all of $\gamma$ and giving a bound of $\rho^{|\gamma|/20 + 1}$ on the event of interest. 

To obtain the stronger statement in the lemma, encircling $\gamma(1)$ involves tracing out a different slightly longer curve and using smaller balls of radius $r/10$. The conditions $r/10 > 10$ and $r/10 > R$ result in the stated bounds on $r$.\hfill $\qed$

\vspace{3pt} 

\noindent While this key lemma appears in previous work, significant additional effort was required to apply it toward our new results.

\subsection{Grid-like Graphs and Wired Duals}

A \emph{grid-like graph} $H$ is a finite, simply connected subset of a grid-like lattice $\Lambda$. By {\it simply connected}, we mean that $H$ has a connected boundary, that is, the set of all vertices in dual graph $\Lambda^*$ corresponding to faces of $\Lambda$ incident on both vertices of $H$ and vertices of $\Lambda \setminus H$ is connected. Informally, this means that $H$ has no holes, which pose a problem for our runtime analysis, though not for the correctness of our algorithm.  

\begin{figure}

\begin{centering}

\includegraphics[scale = 0.65]{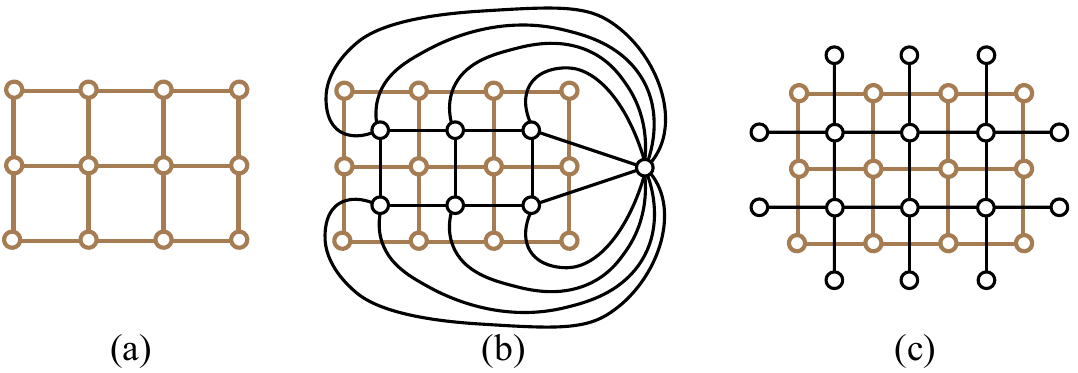}

\end{centering}

    \caption{ (a) A grid graph $H \subseteq \mathbb{Z}^2$. (b) $H$'s dual graph $H^*$. (c) $H$'s wired dual $\wH$.}
    \label{figWiredDual}
\end{figure}

Rather than considering $H$'s dual graph $H^*$, where $H$'s outer face is contracted to a single vertex, at times it is more helpful to consider $H$'s {\it wired dual} $\wH$ (see Fig.~\ref{figWiredDual}). All graphs $H$ we consider are subgraphs of an infinite lattice $\Lambda$, and $\wH$ contains distinct vertices for all faces of $\Lambda$ incident on an edge whose endpoints are both in $H$. We refer to such vertices as the {\it external} vertices of $\wH$. $\wH$ also contains all edges dual to an edge in $H$.  The reason we use a wired dual is that the dual random walks we consider are in fact random walks in $\Lambda^*$, which might leave $H^*$ and then later return along a different edge. Note $\wH$ is a subgraph of $\Lambda^*$, while $H^*$ is~not.

\subsection{Wilson's Algorithm and Modifications}\label{secWilson}

Wilson's algorithm~\cite{wilson} is a well-known method for generating perfectly random spanning trees in a graph $G$. An arbitrary vertex $r$ of $G$ is designated as the {\it root}, and  initially $T = \{r\}$. While $T$ is not yet a spanning tree, a loop-erased random walk starting from an arbitrary vertex $v \notin T$ proceeds until it reaches $T$; when it does, this entire path is added to $T$. Surprisingly, for any choice of $r$ and any choices of starting points $v \notin T$,
this produces a uniformly random spanning tree. The power to choose these vertices makes Wilson's algorithm particularly useful for a variety of goals, and is key to our analysis.  

Let $H$ be the graph of two recently-merged districts that we wish to split. We run Wilson's algorithm on $H^*$ with the following two modifications. First, we run Wilson's algorithm independently on each 2-connected component of $H^*$. A 2-connected component of $H^*$ necessarily has all corresponding faces of $H$ at the same depth, and this is the reason for our definition of depth: for all connected vertices of $H^*$ at the same depth $d$, we can run Wilson's algorithm treating the single face at depth $d-1$ as the root. This ensures our analysis need only consider 2-connected components of $H^*$, simplifying proofs and avoiding issues arising from non-uniqueness of embeddings.    

Second, we carefully select the starting points for each subsequent loop-erased random walk. We do this using random walks in $\Lambda^*$, which might leave $\wH$. After initially starting a loop-erased random walk in $\wH$ at a point $s$ chosen by consideration of a planar separator, rather than stopping after the random walk reaches a vertex already in the tree (and the resulting loop-erased random walk is added to the tree), the random walk continues until reaching a vertex not yet in the tree. This vertex becomes the starting point for the next loop-erased random walk. We label the loop-erased portions (which are added to the tree) of this long random walk as $P_1$, $P_2$, $\ldots$, $P_\ell$. 
We also set step limits on each such {\it stage}, so only continue this process if we have not yet performed~$t$ total random walk steps, for a carefully chosen $t$. These modifications let us apply Lemma~\ref{lem:curveswalk} to one long random walk rather than several short ones.  They also ensure (after $t$ steps) that the whole process resets, at another vertex chosen according to a planar separator.

\section{Overview of Algorithm and Analysis} \label{secProofOverview}

The full definition and analysis of our algorithm is deferred to Sec.~\ref{secAlgorithm}.
At a high level, we run Wilson's algorithm in the dual graph $H^*$, with $T^*$ initially containing only the vertex representing $H$'s outer face. 
For every path in $H^*$ added to $T^*$, the corresponding edges in $H$ cannot be part of the primal spanning tree that will eventually be dual to the spanning tree produced in $H^*$.  Thus, we view the addition of a path to $T^*$ in $H^*$ equivalently as the {\it deletion} of edges in $H$ until ultimately a spanning tree is produced.  

During this process, eventually some edges of $H$ that have not been removed become {\it bridges}, edges whose removal would disconnect $H$. Such a bridge {\it must} be in the final spanning tree of $H$. Suppose a sequence of edge removals in $H$ (as paths are added to $T^*$ in $H^*$) produces a bridge $e$ with $3n/4$ vertices on one side and $n/4$ vertices on the other side. If the final resulting spanning tree of $H$ were to be splittable, the potential split edge {\it must} be in the part with $3n/4$ vertices. How the spanning tree is completed on the side of $e$ with $n/4$ vertices is irrelevant to deciding splittability.  Using this observation, we can start our next loop-erased random walk from a face on the side of $e$ with $3n/4$ vertices, and we never need to return to the side of $e$ with $n/4$ vertices.

This insight forms the foundation of our divide-and-conquer algorithm.  We will use Wilson's algorithm as if we are sampling a random spanning tree $T^*$ of the dual and thus a corresponding primal tree $T$ of $H$. However, we won't generate all of $T^*$, only as much as needed to decide if $T$ is splittable, and if so, where.  By a careful choice of starting points for our loop-erased random walks in $H^*$, we show that after $O(n)$ time, our algorithm has produced bridge(s) in $H$ that collectively have a constant fraction of vertices on each side. We then recurse only on the larger side of the bridge(s). Overall, this produces a runtime of $O(n)$.

\begin{figure}

\ipncm{.38}{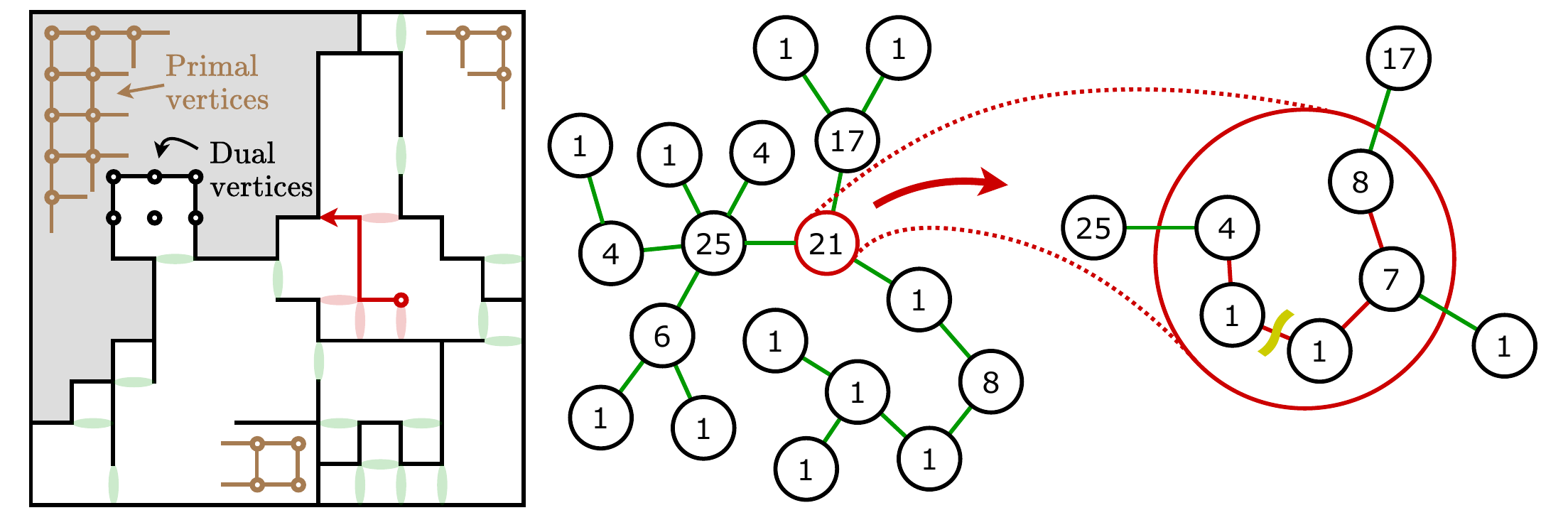}{\label{figRegionTree} Illustration of region trees while running ReCom on a $12 \times 12$ grid graph with $k = 3$ districts and $q = 0$ (exact balance). Left: A given state of running Wilson's algorithm on the dual graph $H^*$, where $H$ is the graph of all white grid cells (the grey cells represent the district not involved in the recombination). A few of the primal and dual vertices are highlighted arbitrarily. Middle: The associated region tree with weights labeled on the vertices. Right: The refinement to the region tree made after running the random walk indicated by the red arrow on the left panel.\vspace{-.8cm}}
\end{figure}

To decide which parts of $T^*$ we need to generate, we maintain a tree-like data structure of all regions induced by the bridge tree decomposition of the primal graph. We call this the {\it region tree}. We also store at each vertex of the region tree the total weight (number of vertices) of each region. See Fig.~\ref{figRegionTree} for an example of an intermediate state of our algorithm and its corresponding region tree. The region tree is updated as dual edges are added to $T^*$ and the corresponding primal edges are deleted.

It is straightforward to see every region tree has a {\it center}, an edge or vertex whose removal separates the tree into subtrees of weight at most half the tree's total weight.  Furthermore, if the region tree has an edge-center, the tree will be splittable, while if the region tree has only a vertex center of weight~1 (an {\it atomic} vertex center), the tree will not be splittable. If the region tree has a vertex center of weight more than $1$, it is in that region that we recurse and begin our next loop-erased random walk per Wilson's algorithm. This results in further refinement of the region tree, as shown in Fig.~\ref{figRegionTree} (right). This process is formally stated in Algorithm~\ref{algMain}.  

\begin{algorithm}[h!]
	\caption{\label{algMain}Samples a tree-weighted balanced partition.}
    \KwIn{A grid-like graph $H$ in a grid-like lattice lattice $\Lambda$, with duals $H^*$ and $\Lambda^*$ and wired dual $\wH$, and an oracle $Q$ for determining pairs $(s,t)$ where $s$ is a random walk starting point and $t$ is a minimum number of steps $t$ to do the random walk for.
    }
    \KwOut{The partition induced by splitting a uniformly random spanning tree of $H$ into two exactly balanced pieces, or $\bot$ if the tree is not splittable.}
    
    $T^* \gets$ the graph containing the outer face of $H^*$ as the only vertex\;\label{linInitializeTStar}
    
    $R(T^*) \gets$ region tree of $T^*$\;
	\While{\texttt{true}}
	{
        \label{linFindEdgeCenter}\If{there exists an edge center $e$ of $R(T^*)$}
        {
            $S_1, S_2 \gets$ sets of regions on either side of $e$ in $R(T^*)$\;
            \KwRet{$(\bigcup_{A \in S_1} \txt{primal vertices in }A,\ \bigcup_{A \in S_2} \txt{primal vertices in }A)$}\;
        }
        \Else
        {
    		$A \gets$ vertex center of $R(T^*)$\;\label{linFindVertexCenter}
            \If{$A$ is atomic}
            {
                \KwRet{$\bot$}
            }
            \While{some region in the refinement of $A$ has weight $> \frac34$ the weight of $A$\label{linFindSeparator}}
            {
                $(s, t) \gets Q(A)$\;\label{linQuery}
                $P_1, P_2, \dots, P_\ell \gets$ Modified Wilson walk from $s$ in $\Lambda^*$ for at least $t$ steps\;\label{linWilsonPhase2}
                \For{$i \in \{1, 2, \dots, \ell\}$}
                {
                    add vertices and edges in $P_i$ to $T^*$\;
                    update $R(T^*)$ by adding edges in $P_i$ and any new doors from $P_i$\;
                }
            }
        }
	}
\end{algorithm}

Much of our work goes into showing there exist starting points for our loop-erased random walks such that, in $O(n)$ time, there is a bridge (or bridges) that collectively separate a constant fraction of vertices from the rest of what remains in $H$. In Algorithm~\ref{algMain}, these starting points $s$ are simply given by an oracle $Q$. Our method for choosing starting points $s$ as well as stage length $t$ is given in a separate Algorithm~\ref{algQ}, stated in Sec.~\ref{subAlgorithmRuntime}. 
A key ingredient in Algorithm~\ref{algQ} is a specific {\it planar separator theorem} (Theorem~\ref{thmPlanarSep}), which states there must be a cycle $c$ in $H^*$ of length $O(\sqrt{n})$ such that there are at most $2/3$ of $H$'s vertices on one side and at most $2/3$ of $H$'s vertices on the other side of the cycle. We show how to pick starting points in $H^*$ for our loop-erased random walks that are near this planar separator $c$, and use Lemma~\ref{lem:curveswalk} to show the subsequent loop-erased random walk approximately traces out~$c$. Note $c$ may get very close to the boundary of $\wH$, and so random walks approximating $c$ might leave and re-enter $\wH$ several times. This is why we use Modified Wilson's algorithm, rather than Wilson's algorithm, to allow for the possibility that a random walk approximately tracing out a planar separator might leave and reenter $\wH$, but nevertheless remain close to the separator.

Some edges will be missing from the route we trace out in $H^*$ that approximates the planar separator $c$. These missing edges will correspond to bridges in $H$ with the desired property: Because the planar separator has at most $2/3$ of the vertices of $H$ on each side of it, the same is true of these bridges (up to some small error as the route we trace only approximates the planar separator).  This gives the desired recurrence. 
Because our planar separator theorem only applies to 2-connected graphs, if a region $A$ we are trying to refine is not 2-vertex connected, we can consider the 2-connected components of $A$ separately and will only need to recurse on at most one of them.

A main difficulty is that this planar separator is itself a cycle. This is problematic as the random walks in Wilson's algorithm erase loops. It's also possible that the planar separator might come arbitrarily close to itself. Because our random walks only approximately trace out the separator, near-loops of this planar separator could result in large sections of our random walk being unintentionally erased. We therefore trace out the planar separator in an careful way to ensure that doesn't happen. We do this by breaking the separator into smaller segments, and tracing each in a backwards direction, towards what has already been traced out. We prove any path can be broken up into a bounded number of segments so that each one is surrounded by a small simply connected buffer region. Using topological arguments involving homotopies of the embeddings of paths in the plane, we ensure these backward-oriented segments do not erase any critical loops. The lengths of these segments inform the choice of parameter $t$, which gives how long a stage of Modified Wilson's algorithm must last: Go at least until the random walk reaches a previously-constructed segment of the separator.

In Sec.~\ref{subAlgorithmDefinition}, we formally define region trees, prove several results related to region trees and spittability, and prove Algorithm~\ref{algMain} is correct. In Sec.~\ref{subAlgorithmRuntime} we provide an implementation of its key subroutine $Q(A)$ to choose a random walk starting point from a region $A$ as Algorithm~\ref{algQ}. Our main result (Theorem~\ref{thmExact2balanced}) follows from combining the following two theorems:

\begin{restatable}{theorem}{thmCorrect}\label{thmCorrect}
    Algorithm \ref{algMain} correctly simulates the following process: Sample a uniformly random spanning tree $T$ of $H$, and output the unique partition into balanced subtrees of $T$ if it exists, otherwise output $\bot$.
\end{restatable}

\begin{restatable}{theorem}{thmMainRuntimeBound}\label{thmMainRuntimeBound}
    Algorithm~\ref{algMain} runs in expected linear time when using policy $Q(A)$ from Algorithm~\ref{algQ}.
\end{restatable}

\noindent Finally, in Sec.~\ref{subTreeSamplingSpeedup} and Sec.~\ref{subApproximateBalance} we consider respective extensions to faster tree sampling and approximately balanced partition sampling.

\section{Algorithm}\label{secAlgorithm}

We now present our algorithm. In Sec.~\ref{secWilsonApp}, we provide more details on the modifications of Wilson's algorithm we use. In Sec.~\ref{subAlgorithmDefinition} we formally define all the data structures and other concepts used in Algorithm~\ref{algMain}. The algorithm is not fully-specified, requiring a policy for choosing the starting parameters $(s,t)$ for each stage of our Modified Wilson's algorithm, where $s$ is the starting dual vertex and $t$ is the minimum number of steps in the stage. We prove that it correctly samples tree-weighted partitions for any choice of these starting points. In Sec.~\ref{subAlgorithmRuntime} we prove that there exist a sequence of starting parameters and step numbers guaranteeing expected runtime $O(n)$. These can be computed efficiently without affecting the $O(n)$ runtime. In Sec.~\ref{subTreeSamplingSpeedup} we give a variant of our algorithm which takes longer time $O(n \log n)$, but works with arbitrary starting points and also produces the entire uniformly random spanning tree. Finally, in Sec.~\ref{subApproximateBalance} we extend to the case of approximate balance.

\subsection{Modifications of Wilson's Algorithm}\label{secWilsonApp}

As briefly outlined in Sec.~\ref{secWilson}, we slightly modify Wilson's algorithm for our use. 
Let $H$ be the graph of the two recently-merged districts that we wish to split. We will run Wilson's algorithm on $H^*$ with the following two modifications. 

\textbf{Modification 1: Restrict to certain 2-connected components.}
Suppose that $H^*$ has a cut vertex; that is, a vertex whose removal disconnects the graph into multiple nonempty connected components $C_1, C_2, \dots, C_\ell$. Choosing a random spanning tree of $H^*$ is thus equivalent to independently choosing a random spanning tree of each $C_i$. So instead of running Wilson's algorithm on the entire graph $H^*$ at once, we may choose to run it separately on each $C_i$, with the provision that random walks are never allowed to cross into any other $C_j$. We choose to do this whenever there are cut vertices in $H^*$ corresponding to faces in the embedding of $H$ that enclose other faces.

Formally, we implement separate random walks based on face depth (defined in Sec.~\ref{sec:dual}). When running a random walk from a (non-outer) face of depth $d$, we will consider the (unique) face of depth $d - 1$ to be already part of the dual tree $T^*$ and never step into a face of depth $d + 1$. For example, in the graph from Fig.~\ref{figDepth}, when running a random walk from the face of depth 3, we will choose the enclosing face of depth 2 as the root, implying the walk will end after a single step that passes through one of the three dual edges. (In terms of the primal spanning tree, this corresponds to choosing which of the three edges on the boundary of the face of depth 3 to delete.) Note that because the spanning tree at each depth is independent of the spanning trees at lower and higher depth (separated by the dual cut-vertex which is the primal surrounding face), we can consider the depths in any order. 

We note that Modification 1 is irrelevant in the case when $H$ is a simply-connected grid subgraph, since all faces aside from the outer face have depth 1. It is only necessary in graphs like Fig.~\ref{figDepth}, where the primal graph is not 2-connected, and moreover when the primal graph has 2-connected components that are separated from the outer face.

\textbf{Modification 2: Using extra walks in $\Lambda^*$ to find starting points for actual walks.}
We construct a spanning tree in $H^*$, but do so by running random walks in $\Lambda^*$ Each {\it stage} of Wilson's algorithm begins with a pair $(s,t)$, where $s$ is the starting point (a vertex of $\Lambda^*$) and $t$ is a minimum number of steps. The stage conducts a loop-erased random walk in $\Lambda^*$ beginning at $s$.  When it reaches a dual vertex already in $T^*$ or reaches an external vertex of $\wH$, the resulting path is added to $T^*$, just as in Wilson's algorithm; if the path ends at an external vertex of $\wH$, this path is attached to the root. If the walk has not yet performed $t$ steps and $T^*$ is not complete, the random walk continues in $\Lambda^*$.  Once the walk reaches a non-external vertex $s'$ in $\wH$ that is not yet in $T^*$, the process repeats, with $s'$ as the new starting point and a minumum number of steps that is $t$ minus the number of steps already performed in the stage. 

Thus, there are alternating phases within each stage: a loop-erased random walk in $\wH$ until $T^*$ is reached, and then a random walk in $\Lambda^*$ (possibly going outside of $\wH$) until a non-external vertex of $\wH \setminus T^*$ is reached, which serves as the new starting point for the next loop-erased random walk. Each time a path is added to $T^*$, we check whether $t$ total random walk steps have occurred, and end the stage if so, otherwise we continue. Essentially, within each stage, we are running Wilson's algorithm but using a random walk in $\Lambda^*$ to choose the starting point of the next loop-erased walk.  

Later, we will use the notation $P_1, P_2, \ldots , P_\ell$ to denote the $\ell$ paths added to $T^*$ during this stage. Sometimes we will consider running Wilson's algorithm in $H^*$ and at other times we will consider running Modified Wilson's algorithm in $\wH$, depending on which is more convenient.

\subsection{Algorithm Details and Correctness}\label{subAlgorithmDefinition}

In this section, we fully define the data structures and other concepts underlying Algorithm~\ref{algMain}, and prove Algorithm~\ref{algMain} is correct. 

In what follows, we let $H$ be the graph of the pair of merged districts, so that our objective is to output a random 2-partition of $H$. At a very high level, our algorithm is running Modified Wilson's algorithm on the wired dual graph $\wH$, but terminating before the entire spanning tree is drawn. All of the work goes into choosing clever starting points for each loop-erased random walk to minimize the expected runtime and provide the information we need to decide the output.

Between these random walks, the state of the algorithm is fully specified by a tree $T^*$ in $H^*$ (not necessarily a spanning tree). We will augment this information with extra data about the planar embedding, in a data structure we call the \emph{region tree}. For any (not necessarily spanning) tree $T^*$ in $H^*$, define $H(T^*) = (V(H), E(H) \setminus \{e \suchthat e^* \in T^*\})$ to be (primal) $H$ with all edges in $T^*$ removed. Graph-theoretically, the region tree $R(T^*)$ is the bridge tree decomposition, also called the bridge-block decomposition~\cite{BridgeTreeDecomposition}, of $H(T^*)$. To form the region-tree, each 2-edge connected component of $H(T^*)$ is contracted to a single vertex whose weight is the number of contracted vertices. The edges between these vertices are all edges that were bridges in $H(T^*)$. See Fig.~\ref{figRegionTree} for an example. We refer to the vertices of $R(T^*)$ as \emph{regions} and the edges as \emph{bridges}. For any region $A$, we overload notation and define $H(A)$ to be the 2-edge connected component of $H(T^*)$ represented by $A$.

For any region tree $R$, we will also store the following additional information:
\begin{itemize}\setlength{\itemsep}{0pt}
    \item To each region $A \in V(R)$, we will store the subgraph of associated vertices in $H$, along with its total weight (that is, its total number of vertices). If the graph consists of a single vertex, we call $A$ \emph{atomic}.
    \item To each bridge in $E(R)$ we will store the associated bridge edge in $H$. We refer to a bridge's dual edge in $H^*$ as a \emph{door} (denoted as green ovals in Fig.~\ref{figRegionTree}).
    \item We keep track of the cyclic order of the the edges incident to each vertex of $R$ induced by the order of the doors in the embedding. (This is optional, only needed for a practical speedup.)
\end{itemize}

\noindent Our next lemma characterizes changes to the region tree after an iteration of Wilson's algorithm.

\begin{lemma}\label{lemRefinement}
    Let $R_1=R(T^*_1)$ and $R_2=R(T^*_2)$ be the region trees immediately before and after a completed loop-erased random walk of Modified Wilson's Algorithm starting from dual vertex $v^*\notin T^*_1$. Then $R_2$ is obtained from $R_1$ by replacing the region $A$ containing the boundary vertices of $v^*$ with a tree linking the neighbors of $A$ together.
\end{lemma}

\begin{proof}

We first claim that the walk from $v^*$ will not end up altering any other region $B \neq A$. By definition, there is some primal bridge edge $e$ separating $A$ from $B$. Since $e$ is a bridge, its dual edge $e^*$ in $H(T_1^*)$ is a self-loop on a face $f$ that separates the two ends of $e$. There are thus two cases, depending on which region is on which side. If $A$ is on the outside of $f$ and $B$ is on the inside, Modification 1 ensures the random walk will never enter $B$, because passing through $f$ requires increasing the depth. Similarly, if $A$ is on the inside of $f$ and $B$ is on the outside, Modification 1 ensures the random walk will stop as soon as it enters $f$, as the depth decreases. Thus, in neither case is the random walk able to pass from $A$ to $B$.

Looking more closely at what happens within $A$, let $C = H(A)$ be the 2-edge connected component of $H(T_1^*)$ that was contracted to form region $A \in V(R_1)$. Note $H(T_2^*)$ differs from $H(T_1^*)$ by the {\it deletion} of the edges in $H$ that are dual to path $P^*$; call this set of edges $P$.  Vertices in $C$ that were previously 2-edge connected may now no longer be 2-edge connected in the graph $C \setminus P = (V(C), E(C) \setminus P)$. This means the region-tree decomposition now turns $C$ into the region tree $R(C \setminus P)$, and this tree replaces $A$ in $R_2 = R(T_2^*)$.  As all bridges adjacent to $A$ are still adjacent to some vertex of $C \setminus P$ as only edges have been removed, $R(C \setminus P)$ must necessarily link the neighbors of $A$ together. 
\end{proof}

We denote the operation transforming $R_1$ into $R_2$ from Lemma~\ref{lemRefinement} as \emph{refining} $A$. An example of refinement is shown in Fig.~\ref{figRegionTree}, where the vertex labeled as 21 is refined into a tree on 5 vertices.  A consequence of this lemma is that Wilson's algorithm is finished once we have recursively refined all regions until they represent individual vertices of $H$.

\begin{lemma}\label{lemRefineST}
    Let $R$ be a region tree. Repeatedly refining $R$ will result in every vertex being atomic and $R$ becoming a spanning tree of $H$.
\end{lemma}

\begin{proof}
    This follows from the fact that Wilson's algorithm on $H^*$ ultimately produces a spanning tree $T^*$ on $H^*$, which bijectively corresponds to a spanning tree $S$ on $H$. By Lemma~\ref{lemRefinement}, each loop-erased random walk in $H^*$ that is added to $T^*$ corresponds to removing edges from some region of $R$, possibly resulting in the creation of new regions if vertices that were previously 2-edge connected no longer are. Eventually, the edges whose duals have not been added to $T^*$ form a spanning tree $S$ of $H$ by Lemma~\ref{lemDuality}.  These are the edges that are considered when forming $R(T^*)$, and as $H(T^*)$ is now a tree, no vertices are two edge-connected. Thus all vertices of $R(T^*)$ are atomic, and $H(T^*) = R(T^*) = S$.     
\end{proof}

The efficiency savings of our new algorithm will come from the fact that we do not need to refine all regions to determine whether the final resulting tree will be splittable or not. To decide which regions to refine, our recursive algorithm begins by applying the following lemma to the current region tree.

\begin{restatable}{lemma}{lemTreeCenter}\label{lemTreeCenter}
    Every vertex-weighted tree has a center, which is either an edge or a vertex whose removal separates the tree into subtrees of weight at most half the total weight. Moreover, it is possible to find an edge center if one exists, otherwise a vertex center, in linear time (in the number of vertices in the tree). When there is no edge center, the vertex center is unique.
\end{restatable}

A variant of this result is well-known for edge-weighted trees~\cite{Megiddo}, and easily extends to vertex-weighted trees. For completeness, we include a proof in Appendix~\ref{appTreeCenter}. The following lemma shows how the refinement operation affects graph centers. 

\begin{lemma}\label{lemRefinementPreservesCenter}
Suppose region tree $R_1$ has unique vertex center $A_c$. Then if region tree $R_2$ is obtained from $R_1$ by refining $A_c$, the new center is one of the new regions or bridges added to~$R_2$.
\end{lemma}

\begin{proof}
    As $R_1$ has a unique vertex center, no edges or vertices except $A_c$ are centers.  This means for any edge or vertex that is not $A_c$, removing it from $R_1$ results in at least one subtree of total weight greater than half the total weight. 

    Note refinement, as defined by Lemma~\ref{lemRefinement}, preserves total weight, as the weights of vertices in any region tree simply count the number of vertices contracted to form each region. This means that also in $R_2$, for any edge or vertex that is not one of the new edges or vertices introduced in the refinement, removing it from $R_2$ results in at least one subtree of total weight greater than half the total weight. It follows that no such vertex or edge can be a center of $R_2$.  Since Lemma~\ref{lemTreeCenter} states every vertex-weighted tree has a at least one center, this center must one of the new regions or bridges introduced in the refinement. 
\end{proof}

\noindent We now prove our key lemmas relating region trees and centers to splittability. We say that a tree is $q$-balanced if there is an edge whose removal from the tree yields a $q$-balanced partition. In this case exact splittability is equivalent to $0$-balance. 

The following is immediate from the definitions:
\begin{lemma}\label{lemCenterSplittability1}
    A spanning tree $T$ is 0-balanced if and only if has an edge center.
\end{lemma}.

Thus, our goal for determining splittability to is to refine the region tree enough to determine whether or the tree has an edge center.

\begin{lemma}\label{lemCenterSplittability2}
    If $R(T^*)$ has no edge center but has an atomic vertex center, and $S\propto R(T^*)$, then $S$ is not 0-balanced.
\end{lemma}
\begin{proof}
    Since $R(T^*)$ has no edge center the atomic vertex center of $R(T^*)$ is unique.    Suppose $S$ has an edge center $e$. Then $e$ must lie entirely in some the graph $H(A)$ for some region $A\in V(R)$; otherwise, $e$ corresponds to an edge center of $R(T^*)$, contradicting our hypothesis.  But then $A$ is a vertex-center for $R(T^*)$, and $A$ is not atomic, so the atomic vertex center was not the unique vertex center, a contradiction.
\end{proof}

\begin{lemma}\label{lemCenterSplittability}
    If a region tree $R(T^*)$ has an edge center, then any spanning tree $S$ resulting from refinements of $R(T^*)$ is 0-balanced.  If a region tree $R(T^*)$ has no edge center but has an atomic vertex center, then any spanning tree $S$ resulting from refinements of $R(T^*)$ is not 0-balanced. 
\end{lemma}

\begin{proof}
First, let $R(T^*)$ be a region tree with an edge center $e$.  This means that removing $e$ from $R(T^*)$ results in two subtrees each of weight at most half the total weight.  Since $R(T^*)$ is vertex-weighted, all weight thus must be in one of these two subtrees, meaning each subtree of $R(T^*)$ must have exactly half the total weight. Any refinements to $R(T^*)$ will not change this, as refinements only affect vertices and will keep the weight in each subtree the same.  When $R(T^*)$ is repeatedly refined, by Lemma~\ref{lemRefineST} the result is a spanning tree $S$ of $H$.  As removing $e$ from $S$ results in two subtrees with the same total weight, i.e. the same  number of vertices, this means $S$ is 0-balanced. 

Next, suppose $R(T^*)$ has no edges but has an atomic vertex $A_c$ as its center. This means that removing $A_c$ from $R(T^*)$ results in subtrees that each have total weight at most half the total weight, where the total weight is the number of vertices in $H$.  Consider any such subtree $T$ obtained from $R(T^*)$ by deleting $A_c$.  This subtree has less than half of the total weight, meaning the vertices not in $T$ (which are $A_c$ together with the other subtrees) must have cumulative weight greater than or equal than half the total weight. Additionally, the vertices not in $T$ cannot have cumulative weight exactly half the total weight, as in this case the bridge from $A_c$ to $T$ would be an edge center, and we know $R(T^*)$ has no edge centers.  Therefore the vertices not in $T$ have cumulative weight strictly greater than half the total weight. 

When $R(T^*)$ is repeatedly refined, by Lemma~\ref{lemRefineST} the result is a spanning tree $S$ of $H$. Let $e$ be any edge in $S$ that is a bridge in $T$, has both endpoints in a region of $T$, or is the bridge connecting $T$ to $A_c$. When any such $e$ is removed from $S$, then $A_c$ and all vertices from regions of $R(T^*)$ that are not in $T$ are in the same component.  As these vertices must be  strictly more than half of the total vertices, edge $e$ cannot be a balance edge whose removal leaves $S$ with two components of the same size.  Applying the same argument to the other subtrees obtained from $R(T^*)$ by deleting $A_c$, we see there is no edge in $S$ whose removal produces two components of the same size.  Thus $S$ cannot possibly be 0-balanced. 
\end{proof}

Utilizing the previous lemmas, our main algorithm is presented as Algorithm~\ref{algMain}.

\thmCorrect*

\begin{proof}
Suppose you generate a random spanning tree of $H$ by the following process: while Algorithm~\ref{algMain} has not yet returned a value, begin generating a random spanning tree in the dual graph $H^*$ by choosing the same starting points as Algorithm~\ref{algMain} for each loop-erased random walk and making the same random choices.  Once Algorithm~\ref{algMain} has returned something, complete the dual random spanning tree using any starting points for the remaining loop-erased random walks. Finally, return the spanning tree of $H$ that is dual to the spanning tree your process produced. Because $H$ is produced by a particular implementation of Wilson's Algorithm, it is a uniformly random spanning tree by the correctness of Wilson's Algorithm.

By Lemmas \ref{lemCenterSplittability1} and \ref{lemCenterSplittability2}, this spanning tree is 0-balanced and splits into components $(S_1, S_2)$ if and only if Algorithm~\ref{algMain} returns $(S_1, S_2)$, and this spanning tree is not 0-balanced if and only if Algorithm~\ref{algMain} returns $\bot$. Thus Algorithm~\ref{algMain} correctly simulates the process claimed.   
\end{proof}

Finally, we note one additional speedup that we can apply, which does not affect the asymptotic runtime complexity, but we expect should improve efficiency in practice. The proof is deferred to Appendix~\ref{appSpeedup}.

\begin{restatable}{proposition}{proSpeedup}\label{proSpeedup}
    Let $A$ be a vertex which is a center of region tree $R(T^*)$. If there is a 0-splittable spanning tree $S$ resulting from refinements of $R(T^*)$, then there is a partition of the neighbors of $A$ into two sets $A_1$ and $A_2$, each contiguous in the cyclic order of the embedding, such that the total weight of vertices in the subtrees of each $A_i$ is at most half the total weight.
\end{restatable}

Observe that this partition condition is checkable in linear time in the maximum degree of the region tree. We simply keep track of two pointers moving around the cycle of neighbors. Thus, we may check this condition in place of line \ref{linFindVertexCenter}; if is not satisfied, we may immediately stop because the tree cannot possibly be splittable.

\subsection{A Linear-Time Implementation}\label{subAlgorithmRuntime}

In this section we show that there exists an implementation of oracle $Q$ that makes Algorithm~\ref{algMain} run in linear time.  We begin with some additional preliminaries. 

\subsubsection{Planar Separator Theorem}

The Planar Separator Theorem, first proved by Lipton and Tarjan~\cite{LiptonTarjan}, states that, in any planar graph of $n$ vertices, we can efficiently find a set of $O(\sqrt{n})$ vertices separating the graph into components of size at most $\frac{2n}{3}$. We will need the following variant of this result by Miller~\cite{CycleSeparators1}, where the separator is a cycle and we count the faces on either side of the separator, rather than the vertices. This result requires an additional assumption that the degree of the dual graph is bounded by a constant, $b$.

\begin{theorem}[Planar Separator Theorem \cite{CycleSeparators1}]\label{thmPlanarSep}
    Let $G$ be an embedded 2-connected planar graph with $n$ vertices, at least two faces, and maximum dual degree $b$. There exists a simple cycle in $G$ of size at most $2\sqrt{bn}$ such that at most $\frac23$ of the faces are in the interior and at most $\frac23$ of the faces are in the exterior. Moreover, one can find such a cycle in time $O(n)$.
\end{theorem}

We note that we will apply this theorem to subgraphs of the dual lattice $\Lambda^*$, the faces of which are vertices in the primal lattice $\Lambda$.  For example, this occurs in Line 5 of Algorithm~\ref{algQ}, where we obtain a cycle of faces in $H$ (that is, of vertices in $H^*$) separating a region $A$ of $H$.
That we apply this theorem to the dual is why we assume that the primal vertices have bounded degree in the definition of a grid-like lattice.

An issue is that this theorem only applies to 2-connected planar graphs, and we do not necessarily know that the dual subgraph we apply it to will be 2-connected.  The following definitions and lemmas are necessary to achieve this. 

\begin{definition}
A {\it 2-connected center component} is a 2-connected component of a graph such that if the component is removed, then each remaining component has fewer than $n/2$ vertices. 
\end{definition}

\begin{lemma}
Let $A$ be a region in a region tree that is a vertex center. It contains a 2-connected center component that can be found in time $O(n)$, for $n$ the number of vertices in region $A$. 
\end{lemma}
\begin{proof}
    First, if $H(A)$ is 2-connected, then $H(A)$ itself must be a 2-connected center component because $A$ is a vertex center of the region tree. 

    If $H(A)$ is not 2-connected, decompose it into its 2-connected components. This forms a tree known as the {\it block-cut tree}, which has vertices of two types: block nodes representing 2-connected components with weights consisting of the number of non-cut vertices each contains, and cut nodes representing cut vertices with weight one.  It is straightforward to see (for example, see the proof of Lemma~\ref{lemTreeCenter} in the appendix) that this vertex-weighted tree must have a vertex center $c$.  If $c$ is a block node, that 2-connected block (including its cut vertices) form a 2-connected center component. If $c$ is a cut node, then any component containing $c$ is a 2-connected center component.   As the number of vertices in the block-cut tree is $O(n)$, the desired center can easily be found in $O(n)$ steps using a depth-first search-type approach in the block-cut tree. 
\end{proof}

While the following lemma appears in several sources, we could not locate a proof, and so include a proof here for the sake of completeness.

\begin{lemma}
    Let $S$ be a 2-connected simple graph.  Then its dual $S^*$ is 2-connected. 
\end{lemma}
\begin{proof}
First, because $S$ is connected, $S^*$ must be connected. As $S$ is 2-connected, it has no cut-edges, and thus $S^*$ has no self-loops.

Suppose, for the sake of contradiction, that $S^*$ is not 2-connected.  This means $S^*$ must have a cut vertex $x$. Let $S_1^*$ and $S_2^*$ be two components of $S^*$ produced when $x$ is removed. Vertex $x$ must be connected in $S^*$ to at least one vertex in $S_1^*$ and at least one vertex in $S_2^*$. This means there is at least one face $f$ of $S^*$ incident on vertex $x$ that has, along its boundary, a vertex of $S_1^*$, followed by $x$, followed by a vertex of $S_2^*$.  Call these vertices adjacent to $x$ along the boundary of this face $x_1$ and $x_2$, respectively. 

Continue along the boundary of face $f$ starting with $x$, then $x_1$, and so on. We can't go directly from a vertex of $S_1^*$ to a vertex not in $S_1^*$, because $S_1^*$ is a (maximal) connected component of $S^* \setminus x$. That means this face boundary will need to again visit $x$ before reaching a vertex not in $S_1^*$.  Consider the closed walk in $S^*$ formed by the boundary of $f$ from $x$ to the first time it returns to $x$. Let $C_1$ be the connected component of $S^*$ on the opposite side  of this closed walk from $f$. 
(It must be true that $C_1 = S_1^*$ because $S_1^*$ is connected, but we do not need this fact so we do not prove it). 
 Note $C_1$ cannot be a tree, as face $f$ is a vertex in $S$ and this would mean there would be a self-loop in $S$, a contradiction as $S$ is simple. Therefore $C_1$ must have at least one face, say $f_1$.

 Similarly, we can continue along the boundary face of $f$ starting with $x$ then $x_2$, and find a closed walk separating face $f$ from another face $f_2$. 

 Because of how we defined $f$ and our closed walks, any path in $S$ from the vertex representing $f_1$ to the vertex representing $f_2$ must pass through the vertex representing $f$. This means the vertex representing $f$ in $S$ must be a cut vertex of $S$, a contradiction as $S$ is 2-connected.
\end{proof}

As all lattice-like graphs we consider are simple, these results together imply we can always find a 2-connected primal subgraph on which to recurse. Furthermore, this subgraph's dual must also be 2-connected, meaning our planar separator theorem applies.

\subsubsection{Curves, Loops, and Homotopies}\label{subHomotopies}

Our arguments rely on the topology of the plane, so we will find it useful to work with the continuous curves in $\rr^2$ rather than paths in a graph. A \emph{curve} is continuous map from $[0, 1]$ to the plane. A \emph{loop} is a continuous map from the circle $S^1$ to the plane. For two curves $\gamma$ and $\gamma'$ such that $\gamma(1) = \gamma'(0)$, we write $\gamma * \gamma'$ to denote the concatenation of the two curves, i.e.,
$$(\gamma * \gamma')(s) := \twocases{\txt{if } s \leq \frac12}{\gamma(2s)}{\txt{if } s \geq \frac12}{\gamma'(2s-1)}.$$
We write $\gamma\inv$ to denote the reverse path, i.e., $\gamma\inv(s) := \gamma(1 - s)$.

A \emph{homotopy} between two loops $\alpha$ and $\beta$ is a continuous deformation of one loop into the other; formally, a continuous map $F: S^1 \times [0, 1] \to \rr^2$ such that, for all $s \in S^1$, $F(s, 0) = \alpha(s)$ and $F(s, 1) = \beta(s)$. If the image of $F$ lies within some set $X \subseteq \rr^2$, we say that $F$ is a \emph{homotopy within $X$}, and that $\alpha$ and $\beta$ are \emph{homotopic within $S$}. We will need the following fact about homotopies.

\begin{lemma}\label{lemHomotopySeparation}
    Let $X$ be a compact set in the plane, and let $\alpha$ and $\beta$ be two loops that are homotopic within $X$. Any pair of points outside $X$ are separated by $\alpha$ if and only if they are separated by $\beta$.
\end{lemma}

\begin{proof}
    Let $F$ be a homotopy from $\alpha$ to $\beta$ and let $x$ and $y$ be points that that are separated by $\alpha$. Let $t_0\in [0,1]$ be the supremum of $t$ for which $x$ and $y$ are separated by $F(t)$. If $t_0 = 1$, then $x$ and $y$ are separated by $\beta$. Otherwise, $F(\cdot, t_0)$ is a loop that passes through $x$ or $y$, contradicting the assumption that the image of $F$ is contained in $X$.
\end{proof}

\subsubsection{Oracle $Q$ and Runtime Proof}

Our implementation of oracle $Q$, which chooses a starting dual vertex and minimum runtime for each stage of Modified Wilson's Algorithm, is presented as Algorithm~\ref{algQ}. In what follows, we refer to dual vertices in $\Lambda^*$ as {\it faces} to avoid confusion with the primal vertices of $H$.

\begin{algorithm}[H]
	\caption{\label{algQ}Finds a starting point and minimum runtime for the next random walk, to be used as subroutine $Q$ in Algorithm \ref{algMain}.}
    \KwIn{A region $A$ in a grid-like lattice $\Lambda$ with parameters $b, \rho, R, C, L$, containing $n$ primal vertices}
    \KwOut{A pair $Q(A) = (s, t)$, where $s \in V(\Lambda^*)$ is the starting point and $t \in \mathbb{N}$ is the minimum number of steps for which to run Modified Wilson's algorithm}

    \If{A is not 2-connected\label{lin2conn_cond}}
    { $B \gets$ the 2-connected center component of $A$ \;
    \KwRet{Q(B)} \label{linreturnB}\; }
    $\varepsilon \gets 1/(144Cb)$\label{linSetEpsilon}\;
    $r \gets \varepsilon \sqrt{n}$\label{linDefEpsilon}\;
    \If{$r < \max\{100, 10R\}$}
    {
        \KwRet{$(\textnormal{any face not in the dual tree}, 1)$}\label{linEarlyRReturn}\;
    }
    $c \gets$ deterministicly-chosen simple cycle separator of $H(A)^*$\label{linChooseCycleSeparator} as a cycle/path in $\overline{H(A)^*}$\;
    \If{$c$ does not contain a boundary face of $A$, and this is the first time calling $Q(A)$}
    {
        \KwRet{$(\textnormal{any face in $c$}, 1)$}\label{linEarlyConnectionReturn}\;
    }
    cyclically permute $c$ so that it starts and ends at a boundary face\;
    $v_0 \gets$ endpoint of $c$\;
    $m \gets$ number of faces in $c$\;
    $\gamma \gets$ curve in the plane corresponding to $c$\;
    $s \gets$ $v_0$ with probability $\frac12$, otherwise a uniformly random face from $c$\label{linChooseStartPoint}\;
    $t \gets 4LCr(\abs{\gamma} + r)/\left(\frac14 \cdot \min\left\{1,\frac{r}{6m}\right\} \cdot \rho^{100\abs{\gamma}/r + 301}\right)$\label{linChooseRuntime}\;
    \KwRet{$(s, t)$}\label{linQReturn}\;
\end{algorithm}

We emphasize that the choice of cycle separator in Line~\ref{linChooseCycleSeparator} is the same each time $Q(A)$ is called. Algorithm~\ref{algMain} calls this subroutine repeatedly on the same region $A$, so each time it will find the same path $c$. As edges get added to $T^*$ after each call, some of the edges in $c$ will be contracted in the dual graph. However, this does not change $c$ as a path in the wired dual graph.

It will not be sufficient to simply trace out the separator using Lemma~\ref{lem:curveswalk}, since Wilson's algorithm may end up erasing the entire loop. Instead, we break the separator into smaller pieces, and trace them out one-at-a-time.

For any $r > 0$ and any curve $\gamma$ in the plane, we define the \emph{$r$-tube of $\gamma$} to be the set of points within a distance of $r$ to any point in $\gamma$, i.e.
$$B_r(\gamma) := \{x \in \rr^2 \suchthat d(x, \gamma) \leq r\}.$$
Note that we have the following:
\begin{lemma}\label{lemBallToTube}
  For a grid-like lattice with upper density parameter $C$, there are at most $4C r(|\gamma|+r)$ dual vertices of $\Lambda^*$ and at most $4C r(|\gamma|+r)$ primal vertices of $\Lambda$ in the $r$-tube of $\gamma$.
\end{lemma}
\begin{proof}
    We can cover the $r$-tube of $\gamma$ with $\lceil|\gamma|/r\rceil\leq \gamma/r+1$ balls of radius $2r$.  By the density constraint on grid-like lattice graphs, each such ball contains at most $4Cr^2$ primal vertices and at most $4Cr^2$ dual vertices, and the claimed bound follows.
\end{proof}

\begin{lemma}\label{lemNumberOfSimplyConnectedPieces}
	Let $\gamma$ be a curve in the plane and let $r > 0$. There exists a decomposition $\gamma = \gamma_1 * \gamma_2 * \dots * \gamma_{\ell}$, where $\ell = \frac{\abs{\gamma}}{\pi r} + 1$ such that the $r$-tubes of each $\gamma_i$ are simply-connected.
\end{lemma}
\noindent Here, $*$ indicates concatenation of curves.

\ipncm{1.2}{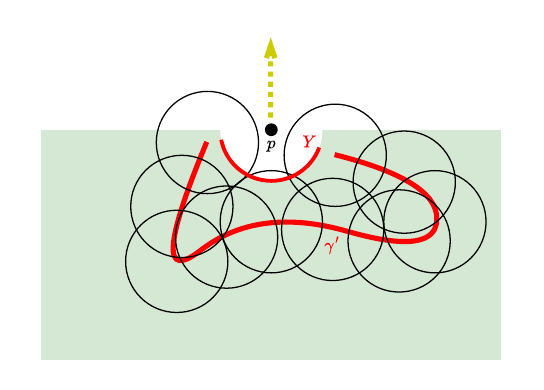}{\label{figRotateAboutP}Illustration accompanying the proof of Lemma~\ref{lemNumberOfSimplyConnectedPieces}. If $Y$ has arc length less than $\pi$, then the $r$-tube of $\gamma'$ does not encircle $p$ as claimed, since there is a ray from $p$ to infinity (the gold dashed line) that does not intersect the $r$-tube (black circles).}
    
\begin{proof}
	It suffices to prove the following: For any curve $\gamma$ and $r > 0$, there is a prefix $\gamma''$ of $\gamma$ (meaning a sub-curve of $\gamma$ sharing a common starting point) of length at least $\min\{\abs{\gamma}, \pi r\}$, such that the $r$-tube of $\gamma''$ is simply-connected. This suffices because we may iteratively apply this claim to bite off $\ell$ pieces of $\gamma$ until the entire curve is covered.
	
	If the $r$-tube of $\gamma$ is simply-connected, we are done, and may simply let $\gamma'' = \gamma$. Otherwise, let $\gamma'$ be any prefix of $\gamma$ such that the interior of the $r$-tube of $\gamma'$ is \emph{not} simply-connected; we claim that $\gamma'$ has length strictly greater than $\pi r$. Pick any point $p$ in an interior hole of the $r$-tube of $\gamma'$. Let $Y$ be the projection of $\gamma'$ toward $p$ to the circle of radius $r$ around $x$. Note that $Y$ is connected since $\gamma$ is, and the length of $Y$ is less than the length of $\gamma'$, since $\gamma'$ lies entirely outside of the circle of radius $r$. We claim that the arc-length of $Y$ is at least $\pi$. Supposing this is not the case, we may re-orient our coordinate system as in Fig.~\ref{figRotateAboutP} so that $p$ is at the origin, and we rotate the curve $\gamma'$ about $p$ so that it lies entirely in the bottom half of the plane, i.e., in
	$$\{(x, y) \in \rr^2 \suchthat y < 0 \txt{ and } x^2 + y^2 > r^2\}$$
	(the green shaded region in Fig.~\ref{figRotateAboutP}). Note that no point in this set is within distance $r$ of any point on the non-negative $y$-axis, including $p$. This means that $p$ is not enclosed by the $r$-tube of $\gamma'$, which is a contradiction. Hence, the arc-length of $Y$ is at least $\pi$, so its actual length is at least $\pi r$. We thus have
	$$\abs{\gamma'} > \abs{Y} \geq \pi r,$$	
	so we may let $\gamma''$ be a prefix of $\gamma'$ of length $\pi r$, which, by the definition of $\gamma'$, will have a simply-connected $r$-tube.
\end{proof}

We next establish our main lower bound on the probability that a sequence of walks of Modified Wilson's algorithm follow the cycle separator closely.

\begin{lemma}\label{lemProbabilityBoundExactBalance}
	Suppose we repeatedly run Modified Wilson's algorithm on a simply-connected region $A$ according to policy $Q(A)$ from Algorithm~\ref{algQ}. Also suppose that $r \geq \max\{100, 10R\}$, so that $Q(A)$ does not return on Line~\ref{linEarlyRReturn}. Let
    \begin{equation}\label{equProbabilityBoundExactBalance}
		B := \min\left\{1,\frac{r}{6m}\right\} \cdot \rho^{100\abs{\gamma}/r + 301}.
	\end{equation}
    After running Modified Wilson's algorithm at least $\ell$ times, where $\ell = \abs{\gamma}/(\pi r)$, plus one additional time in the case where $Q$ first returns on Line~\ref{linEarlyConnectionReturn}, with probability at least $(B/4)^\ell$ there will be some path $c'$ in $A$ such that:
	\begin{enumerate}[(1)]
		
		\item\label{itmProbabilityBoundExactBalanceSmallGaps} Every vertex in $c'$ will have been added to the dual tree.
		
		\item\label{itmProbabilityBoundExactBalanceCycleSeparates} Any pair of dual vertices outside of the $r$-tube of $\gamma$ are separated by $c$ if and only if they are separated by $c'$.
	\end{enumerate}
\end{lemma}

\begin{proof}
    After running the first iteration of Modified Wilson's algorithm, we guarantee that, in all further runs, the separator $c$ will start and end at a boundary face of the region so the algorithm will no longer terminate on Line~\ref{linEarlyConnectionReturn}. Assuming this is case, it remains to show that the remaining $\ell$ walks produce a path $c'$ satisfying both properties.
    
	Write $\gamma = \gamma_1 * \gamma_2 * \dots * \gamma_{\ell}$ as in Lemma~\ref{lemNumberOfSimplyConnectedPieces}. Label the endpoints of each of these paths $p_0, p_1, p_2, \dots, p_\ell = v_0$, so that $\gamma_i$ starts at $p_{i - 1}$ and ends at $p_i$. By property (\ref{itmStrongLatticeSequenceClose}) in the definition of a grid-like lattice, $c$ contains at least $\frac{r}{5} - 1 \geq \frac{r}{6}$ points within radius $\frac{r}{5}$ of $p_i$.  Therefore, with probability at least $\frac12 \cdot \min\{1, \frac{r}{6m}\}$, the $i\tth$ random walk is selected to start from one of these vertices. In the final round, we can additionally assure that the random walk starts from $p_\ell$ itself with at least this probability, since $p_\ell = v_0$ is selected with probability $\frac12$. Next, we have by Lemma~\ref{lem:curveswalk} that, with probability at least $\rho^{100\abs{\gamma_i}/r + 301}$, the walk stays within the $r$-tube of $\gamma_i$ and encircles $p_{i - 1}$ at distance at least $\frac{r}{5}$. So in total, the probability that the $i\tth$ random walk is selected from such a starting point and behaves as in Lemma~\ref{lem:curveswalk} as at least $B/2$. 
    
    We additionally want to know that the number of steps for the random walk to trace out the desired path is not too large.     
    By property (\ref{itmStrongLatticeSequenceEscape}) in the definition of a grid-like lattice, the probability that the random walk takes more than $t = 4LC r (\abs{\gamma} + r)/(B/4)$ steps to escape the $r$-tube of $\gamma$ (which contains at most $4C r (\abs{\gamma} + r)$ dual vertices by Lemma~\ref{lemBallToTube}) is at most $B/4$. This means the probability that the random walk will escape the $r$-tube of $\gamma$ in fewer than $t$ steps is at least $1-B/4$.     
    It follows that with probability at least $B/2 - B/4 = B/4$, the random walk escapes the $r$-tube within $t$ steps \emph{and} behaves as in Lemma~\ref{lem:curveswalk}, implying it completes the encircling of $p_{i - 1}$ in at most $t$ steps.
    
    Multiplying these independent probabilities together, and using the bound $\abs{\gamma_i} \leq \abs{\gamma}$, we conclude that the following holds with probability at least the expression given in Equation (\ref{equProbabilityBoundExactBalance}): For all $i$, the $i\tth$ random walk starts within radius $\frac{r}{5}$ of $p_i$ (starting exactly at $p_\ell$ in the last iteration), stays within the $r$-tube of $\gamma_i$, and encircles $p_{i - 1}$ at radius at least $\frac{r}{5}$.

    Suppose that this happens. We will show that there exists a path $c'$ satisfying both properties. We construct $c'$ inductively, as depicted in Fig.~\ref{figConstructCPrime}. Observe that the first random walk of Modified Wilson's algorithm encircles the boundary vertex $p_0$ before leaving the $r$-tube around $\gamma_1$. Thus, it must cross the boundary at some point $q_0$, so the entire path will be added to the tree. Specifically, we choose $q_0$ to be a crossing on a part of the boundary that is connected to $p_0$ via some path $w_1$ along the boundary and contained within the $r$-tube of $\gamma_1$. The second random walk encircles $p_1$ at a distance of at least $\frac{r}{5}$, so it also encircles any points within $\frac{r}{5}$ of $p_1$, including the starting point of the first random walk. Since this point is part of the tree linking toward the boundary, it must also cross a vertex $q_1$ in the tree, so again, the entire path will be added to the tree. Continuing inductively, we will produce a sequence of dual vertices $q_0, q_1, q_2, \dots, q_\ell = p_\ell$ such that, for each $i$, we have a path from $q_i$ to $q_{i - 1}$ contained within the $r$-tube of $\gamma_i$ of Modified Wilson's algorithm steps. Let $c'_i$ denote the loop erasure of the $i\tth$ run, which will also be contained in the $r$-tube of $\gamma_i$. We define $c'$ to be the concatenation of each of the $c'_i$ paths. We define $\gamma'$ and $\gamma'_i$ to be the corresponding curves in the plane.

    \begin{figure}[t]\centering \includegraphics{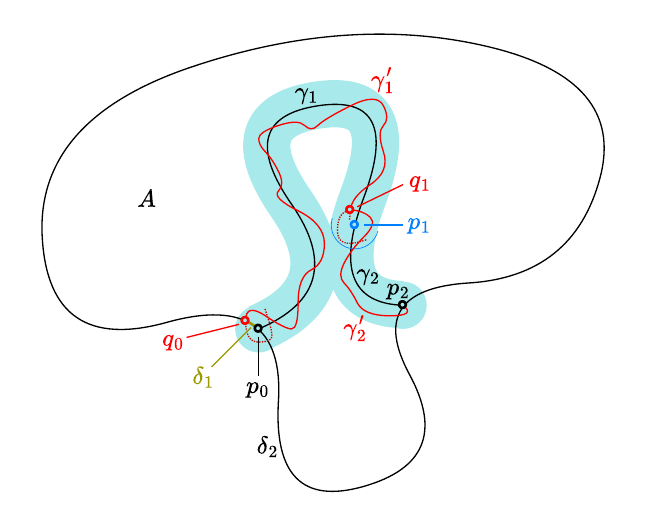}\caption{\label{figConstructCPrime}Illustration of the proof of Lemma~\ref{lemProbabilityBoundExactBalance}. The interior black curve $\gamma$, the path corresponding to the separator $c$. The $r$-tube of $\gamma$ is the shaded region. Since it is not simply connected, it is broken up into $\ell = 2$ pieces according to the proof of Lemma~\ref{lemNumberOfSimplyConnectedPieces}. The red curves show the full trajectories of the two random walks up to the point where they encircle their target point; the dashed parts are excluded from $c'$, so $\gamma'$ consists of only the solid parts of these lines, linking $q_0$ with $p_2 = q_2$.}\end{figure}
    Property~(\ref{itmProbabilityBoundExactBalanceSmallGaps}) holds since $c'_i$ is composed of paths from Wilson's algorithm that terminate at tree vertices. To prove property~(\ref{itmProbabilityBoundExactBalanceCycleSeparates}), let $w_2$ be a path through the dual tree that existed before performing any of the random walks joining the boundary vertices $p_0$ and $p_\ell$. Let $\delta_1$ and $\delta_2$ denote the curves corresponding to $w_1$ and $w_2$, respectively. Let $S$ denote the compact subset of $\rr^2$ that is the union of $\delta_2$ the $r$-tube of $\gamma$. Define the following loops (as concatenations of curves starting and ending at $p_0$):
    \begin{align*}
        \alpha &= \gamma * \delta_2 = \gamma_1 * \gamma_2 * \dots * \gamma_\ell * \delta_2\\
        \beta &= \delta_1 * \gamma' * \delta_2 = \delta_1 * \gamma'_1 * \gamma'_2 * \dots * \gamma'_\ell * \delta_2
    \end{align*}
    We claim that $\alpha$ and $\beta$ are homotopic through $S$. To transition from $\beta$ to $\alpha$, we will send each $q_i$ continuously to $p_i$. The $\delta_1$ segment between $p_0$ and $q_0$ will contract down to a point, and the $\delta_2$ segment between $p_\ell$ and $p_0$ will remain unchanged. Since each $\gamma_i$ and $\gamma'_i$ are contained within the $r$-tube of $\gamma_i$, a simply-connected subset of $S$, we can define the homotopy over these segments with arbitrary trajectories of the endpoints. Thus, applying Lemma~\ref{lemHomotopySeparation}, two points outside of $S$ are separated by $\alpha$ if and only if they are separated by $\beta$. This implies property~(\ref{itmProbabilityBoundExactBalanceCycleSeparates}).
\end{proof}

\noindent We are now ready to prove our main result.

\thmMainRuntimeBound*

\begin{proof}
    Since dual degrees are bounded by a constant $b$, it suffices to bound the runtime as a function of the number of primal vertices, which we will denote by $n$. 
    Because of Lines \ref{lin2conn_cond}--\ref{linreturnB} in Algorithm~\ref{algQ}, we may assume the region $A$ we're considering is 2-connected, and thus the planar separator theorem (Thm.~\ref{thmPlanarSep}) applies.
    Note that the number of dual vertices is at most $bn$. We will show that, for any region $A$ with $n$ primal vertices:
    \begin{enumerate}[(1)]
        \item\label{itmRuntimeBaseCase} If $\varepsilon \sqrt{n} < \max\{100, 10R\}$, then the algorithm terminates in expected constant time.
        \item\label{itmRuntimeRecursiveCase} If $\varepsilon \sqrt{n} \geq \max\{100, 10R\}$, after $O(n)$ time, with probability bounded away from zero, $A$ will be refined into pieces each containing at most $\frac34$ of the vertices of $A$.
    \end{enumerate}
    Note that the latter case will cause the algorithm to escape the while loop on Line~\ref{linFindSeparator}. Since only one of these regions is examined recursively (by Lemma~\ref{lemRefinementPreservesCenter} the new center will indeed be in one of the new regions), the Master Theorem implies the overall expected runtime is $O(n)$.

    To prove (\ref{itmRuntimeBaseCase}), observe that the queries to $Q(A)$ return arbitrary vertices with minimum runtimes of 1. This precisely simulates running ordinary Wilson's algorithm on the dual graph, which will terminate in a constant number of steps, since $n$ is bounded by a constant. If the while loop on Line~\ref{linFindSeparator} is escaped before this happens, the algorithm will only terminate sooner.

    To prove (\ref{itmRuntimeRecursiveCase}), first observe that, since $\gamma$ is the path defined from a separator with at most $2\sqrt{b(bn)} = 2b\sqrt{n}$ dual vertices,
    $$\frac{\abs{\gamma}}{r} \leq \frac{2b\sqrt{n}}{\varepsilon \sqrt{n}} = \frac{2b}{\varepsilon}.$$
    Thus, applying Lemma~\ref{lemProbabilityBoundExactBalance}, after
    $$\left\lceil\frac{2b}{\pi \varepsilon}\right\rceil + 1 = O(1)$$
    iterations, with probability at least
    $$(B/4)^\ell = \left(\frac14 \min\{1, \frac{\varepsilon}{6b}\} \cdot \rho^{100(2b/\varepsilon) + 301}\right)^{2b/(\pi \varepsilon)},$$
    there will exist a path $c'$ satisfying properties (\ref{itmProbabilityBoundExactBalanceSmallGaps}) and (\ref{itmProbabilityBoundExactBalanceCycleSeparates}). Note that this probability is a nonzero constant. For each of the $O(1)$ iterations it takes time
    $$t = 4LC\varepsilon\sqrt{n}(2b\sqrt{n} + \varepsilon\sqrt{n})/\left(\frac14 \cdot \min\{1, \frac{\varepsilon}{6b}\} \cdot \rho^{100(2b/\varepsilon) + 301}\right) = O(n)$$
    to hit the minimum runtime. After that has been achieved, the random succeeds in escaping to the boundary in $2Lbn$ iterations with probability at least $\frac12$ by property (\ref{itmStrongLatticeSequenceEscape}) in the definition of a grid-like lattice. This implies an expected runtime of $O(n)$ overall.
    
    Suppose now that we do obtain such a path $c'$ from Lemma~\ref{lemProbabilityBoundExactBalance}. Note that this path necessarily cuts the active region $A$ into smaller regions, since every edge traversed by Modified Wilson's algorithm is either in the dual tree or between two vertices in the dual tree, in which case it is a door. All that remains to show is that none of the new regions in the refinement of $A$ will have more than $\frac{3n}{4}$ primal vertices. For this it suffices to show that $c'$ separates the primal vertices into components of size at most $\frac{3n}{4}$, since property (\ref{itmProbabilityBoundExactBalanceSmallGaps}) ensures that any gap between consecutive vertices of $c'$ is either in the dual tree or it is a door. Since $c$ separates the primal vertices into components of size at most $\frac{2n}{3}$, the $r$-tube of $\gamma$ also separates the primal vertices into components of size at most $\frac{2n}{3}$. The actual path $c'$ that the random walks generate may fail to separate the components to the same extent; however, property (\ref{itmProbabilityBoundExactBalanceCycleSeparates}) implies that we can bound the error by the number of primal vertices in the $r$-tube of $\gamma$. By Lemma~\ref{lemBallToTube}, there are at most
    $$4Cr(\abs{\gamma} + r) \leq 4C\varepsilon\sqrt{n}\left(2b\sqrt{n} + \varepsilon\sqrt{n}\right) \leq 12C\varepsilon b \cdot n = \frac{n}{12}$$
    such vertices (from the definition of $\varepsilon$ on Line~\ref{linDefEpsilon}). Even if these all were added to the largest component, the largest it could be is $\frac{2n}{3} + \frac{n}{12} = \frac{3n}{4}$.
\end{proof}

\subsection{Exact Uniform Tree Sampling}\label{subTreeSamplingSpeedup}

It is straightforward to extend our algorithm in the previous subsection to sample spanning trees rather than tree-weighted partitions. All one needs to do is recurse on every region in a region tree, rather than only the center region, until all regions in the region tree are atomic, that is, they are single vertices.  By Lemma~\ref{lemRefineST}, at this point we have produced a spanning tree of $H$, as desired. As it takes $O(n)$ work to separate $H$ into pieces each containing at most some constant fraction of the work, and at most $\log(n)$ recursive layers until all regions are atomic, this gives a runtime of $O(n \log n)$ for sampling a spanning tree of a grid-like graph, proving Theorem~\ref{thmST}.

\subsection{Approximate Balance}\label{subApproximateBalance}

We now show how our algorithm can be extended to handle the case of approximate balance. Specifically, fix a function $q(n) = O(n / \log n)$. We show that we can sample $q$-balanced trees on $n$-vertex graphs in $O(n)$ time.

In what follows, fix a vertex-weighted tree $R$ with total weight $n$ (i.e., a given region tree). We say that a vertex or edge is a \emph{$q$-center} if its removal separates the tree into subtrees of weight at most $\frac{n}{2} + q$.

\begin{lemma}\label{lemApproxCenterAlgorithm}
    We can enumerate all vertex and edge $q$-centers in time $O(n)$.
\end{lemma}

\begin{proof}
    We first compute a $0$-center in time $O(n)$ via Lemma~\ref{lemTreeCenter}. The 0-center will always be on the larger side of a $q$-balanced partition. We then root the tree at the 0-center, compute subtree weights using a postorder traversal as in the proof of Lemma~\ref{lemTreeCenter}, and perform a final graph search from the 0-center, keeping track of the total weight accumulated. Any vertices or edges found with weight at most $\frac{n}{2} + q$ are valid $q$-centers.
\end{proof}

\begin{lemma}\label{lemApproxEdgeCenterExists}
    If $R$ has two vertex $q$-centers, then any edge on the path between them is an edge $q$-center.
\end{lemma}

\begin{proof}
    Given two vertex $q$-centers $v_1$ and $v_2$, we may partition the vertices of the tree into three sets, $S_1$, $S_2$, and $S_3$, where $S_3$ contains vertices in between $v_1$ and $v_2$, and for $i \in \{1, 2\}$, $S_i$ contains $v_i$ and all vertices in the sub-tree rooted at $v_i$ after deleting the edges between $v_1$ and $v_2$. Since $v_1$ is a $q$-center, the total weight of $S_2 \cup S_3$ is at most $\frac{n}{2} + q$. Similarly, since $v_2$ is a $q$-center, the total weight of $S_1 \cup S_3$ is at most $\frac{n}{2} + q$. If we remove any edge $e$ in the path between $v_1$ and $v_2$, all vertices on one side are contained in $S_2 \cup S_3$, and all vertices on the other side are contained in $S_1 \cup S_3$; thus, $e$ is a $q$-center.
\end{proof}

\begin{lemma}\label{lemApproxTwoLargeCenters}
    Suppose $q < \frac{n}{6}$. If $R$ has an edge $q$-center $e$, then on either side of $e$, there is a vertex $q$-center $v^*$ such that the total weight of all other vertex $q$-centers on the same side of $e$ as $v^*$ is at most $2q$. Furthermore, $v^*$ can be identified from among the set of vertex $q$-centers in time $O(n)$.
\end{lemma}

\begin{proof}
    Let $C$ be the sub-tree spanned by all $q$-centers. We claim that $C$ is a path. Suppose toward a contradiction that $C$ has some vertex $v$ with at least three incident edges, $e_1$, $e_2$, and $e_3$. By Lemma~\ref{lemApproxEdgeCenterExists}, each $e_i$ is an edge $q$-center. If we were to remove any $e_i$, the side that $v$ is on would have weight at most $\frac{n}{2} + q$, so the side that $v$ is not on has weight at least $\frac{n}{2} - q$. Thus, these three disjoint sub-trees of $C$ have total weight at least
    $$3 \left(\frac{n}{2} - q\right) > 3 \left(\frac{n}{2} - \frac{n}{6}\right) = n,$$
    contradicting the assumption that the total weight is $n$. Hence, $C$ is a path.

    \ipncm{.47}{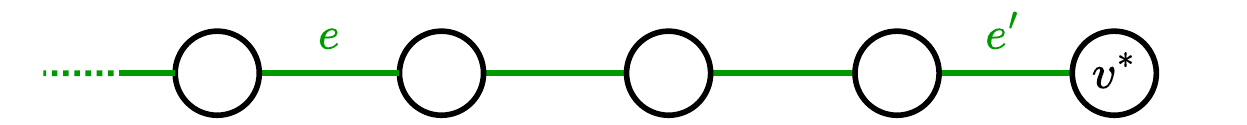}{\label{figCenterPath}Illustration of the path $C$ of all vertex $q$-centers in the proof of Lemma~\ref{lemApproxTwoLargeCenters}.}

    Fix a particular side of the given edge $q$-center $e$, and let $v^*$ be the vertex at the end of the path $C$ on that side, as in Fig.~\ref{figCenterPath}. (Clearly, $v^*$ can be identified in time $O(n)$.) Let $e'$ be the edge incident to $v^*$ in $C$. Again, since $e$ and $e'$ are edge $q$-centers, the weights of all vertices on the side of each of these edges that does not include the other edge are each at least $\frac{n}{2} - q$. Thus, the total weight of all vertices between $e$ and $e'$ in $C$, which includes all vertex $q$-centers except for $v^*$ on the same side of $e$, is at most
    \begin{equation*}
        n - 2\left(\frac{n}{2} - q\right) = 2q. \qedhere
    \end{equation*}
\end{proof}

Our main result for approximate balance is as follows. Note that the output described in the theorem statement is sufficient for its correct use in both ReCom and RevReCom.

\thmApprox*

\begin{proof}
    As in the case of exact balance, we need to refine vertex $q$-centers until only edge $q$-centers and atomic vertex $q$-centers remain. We can still find the $q$-centers efficiently by Lemma~\ref{lemApproxCenterAlgorithm}. The trouble is that there may be more than one that we need to recurse on. As soon as this happens, however, we have by Lemma~\ref{lemApproxEdgeCenterExists} that there is an edge $q$-center $e$. At this point, we run the same algorithm on both sides of $e$, ignoring the other side, but keeping track of its weight.\footnote{Note that we are \emph{not} recursively creating two subproblems multiple times; rather, as soon as we find multiple $q$ centers, we run a different algorithm on each side, which will recurse on only one subproblem each step.} At every step of the recursive algorithm, if there are multiple vertex $q$-centers on the given side of $e$, we apply Lemma~\ref{lemApproxTwoLargeCenters} to identify $v^*$. We then subdivide all other vertex $q$-centers by generating the entire spanning combined tree in time $O(q \log q)$ by Theorem~\ref{thmST}. Note that this step only takes expected time $O(n)$ since $q$ is $O(n / \log n)$. There is only at most one vertex $q$-center to recursively subdivide, namely, $v^*$. Thus, the Master Theorem still applies: After doing expected $O(n)$ work, we recurse on one subproblem of size at most $\frac{3n}{4}$ (the bound from Line~\ref{linFindSeparator} of Algorithm~\ref{algMain}), entailing an overall runtime of $O(n)$.
\end{proof}
\newpage
\section{Computational Experiments}\label{secExperiments}

We now discuss the empirical performance of our bipartitioning procedure and compare it to existing implementations that generate entire spanning trees.

\subsection{Implementation}

We implemented a variant of Algorithm~\ref{algMain} in Python and then optimized in Cython. Below we describe the key aspects in which we deviate from the theory in order to leverage additional efficiencies inherent to our approach and avoid costly (though still linear time) overhead.

\subsubsection{Data Structures}

A guiding principle behind the design of our central data structures is reducing the number of operations that require enumerating over large subgraphs. Crucially, we never maintain a collection of vertices representing a region in the region tree. The underlying graph data structure stores primal and dual nodes and edges with numerous pointers to support operations such as following neighbors, and identifying faces on either side of an edge. A region is defined only by its boundary cycle. Thus, when a region is subdivided into multiple sub-regions after a dual random walk, the running time to update the region tree scales linearly in the \emph{perimeter} rather than number of vertices inside. For a region with $n$ vertices, this represents a speedup from $\Theta(n)$ to as low as $\Theta(n^{5/8})$.\footnote{This is the limiting dimension of loop-erased random walks in $n$-vertex grids, as shown by Kenyon~\cite{Dimension58}.} This is overall asymptotically irrelevant as executing the random walk takes expected $O(n)$ time anyway. However, in the course of implementing and testing our algorithm, it became clear that executing random walks is fast, while updating data structures is the practical bottleneck.

A notable shortcoming of this approach is that we cannot hope to find planar separators without performing expensive graph searches. Instead, we rely on a faster heuristic, described in Sec.~\ref{ssbStartingPoint}. This means that our implementation is not provably linear time.

\subsubsection{Cycle Bases and Populations from the Perimeter}

Algorithm \ref{algMain} relies on testing whether door edges are edge centers, necessitating measurement of region populations on each side of the door edge. This would typically require an enumeration over all nodes contained by a perimeter. To avoid this cost, at preprocessing time we choose a fundamental cycle basis \cite[Sec.~1.9]{GraphTheoryTextbook} for the graph (with respect to an arbitrary spanning tree) and use it to define a set of weights for each dual edge such that the sum of weights around any cycle gives the population of the enclosed primal nodes. We can use additional weight sets in a similar way to seamlessly compute other linear statistics, such as the total expected Democratic/Republican vote share in each district with respect to election data.

\subsubsection{Modified Wilson's Starting Point}\label{ssbStartingPoint}

Eschewing the full planar separator, we use heuristics to choose modified Wilson's starting nodes. We attempted several approaches including selection of a node directly along the perimeter interior, a flood-fill-based centroid selection, uniformly random interior node selection,\footnote{We note that it \emph{is} possible to implement this in time linear in the size of the boundary, without necessitating a flood fill. Nevertheless, the policy still led to a slower overall runtime.} and a short random walk from a random point on the perimeter. The short random walk proved most efficient. The flood-fill and uniform approaches incurred too much overhead, and nodes along the perimeter caused quick collisions of the Wilson's walk to the perimeter boundary which require expensive perimeter walks.

\ipncm{.55}{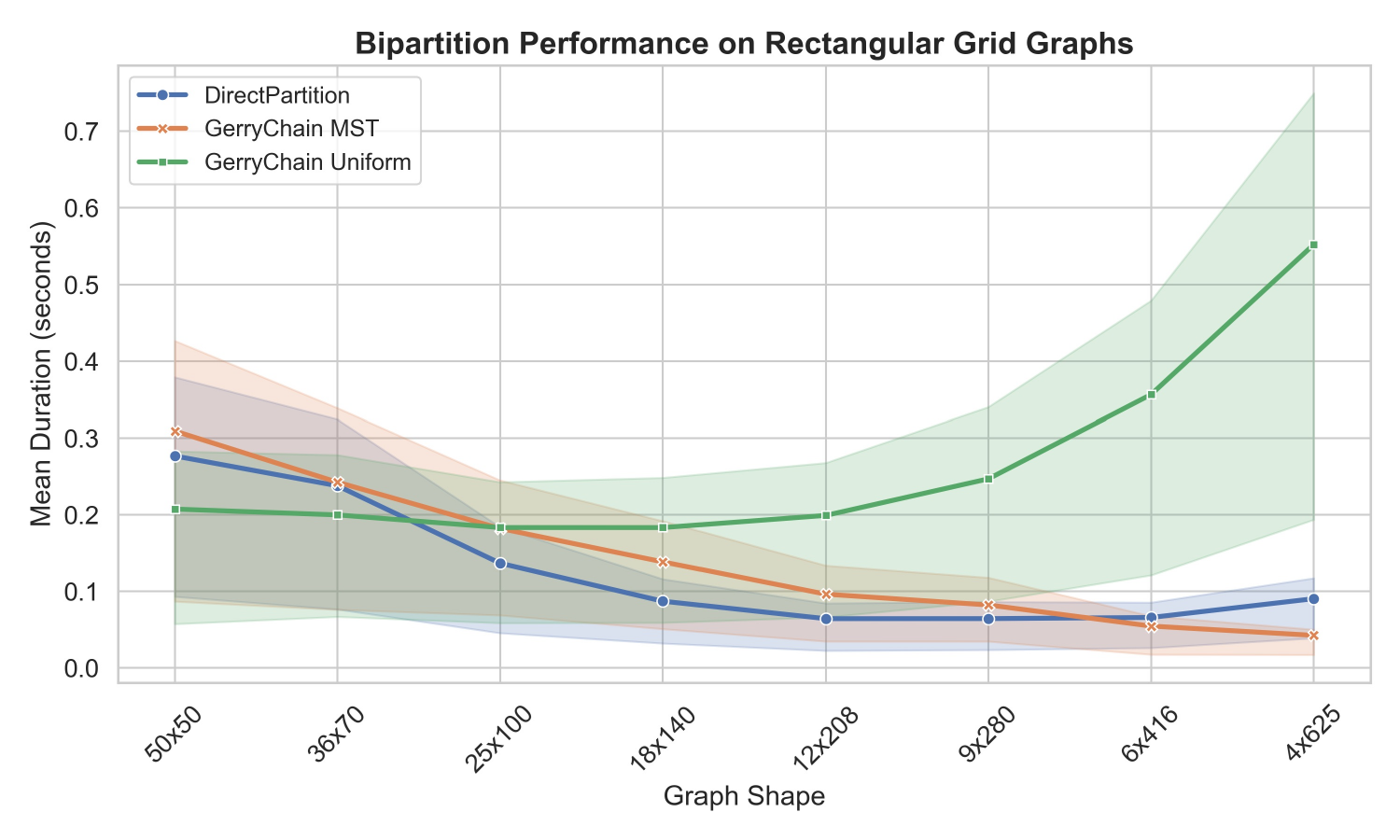}{\label{rectGridGraphs}Mean $2$-split time across grid-graph aspect ratios over $1000$ trials with $25^{th}$ and $75^{th}$ percentile bounds. (Benchmarked on Apple M1 Max.)}

\subsection{Results}

We benchmark our performance against existing methods in the GerryChain repository~\cite{gerrychain}, a leading open-source Python library for running ReCom and related geographic/political data analysis tasks. GerryChain's \texttt{bipartition\_tree} is directly comparable to our new implemtation, which we refer to as DirectPartition. The \texttt{bipartition\_tree} function can either be run on the basis of Wilson's uniform spanning trees (UST) or minimal spanning trees (MST) with respect to random edge weights. We note that comparing DirectPartition to GerryChain's MST algorithm is not a fair comparison, since the distributions over trees are different (and in particular, the theoretical guarantees associated with UST disappear when using MST).

We ran the native Python versions of all three algorithms on grid graphs of various scales, seeking exact bipartitions ($q = 0$). While DirectPartition interacts with fewer nodes overall, it is substantially more complex than a Wilson's-based bipartition and suffers under Python's overhead. As a result, our algorithm underperforms GerryChain UST for square grids of reasonable size (tested up to $4626$ nodes).  (GerryChain MST unexpectedly under-performs both UST and DirectPartition on square grids.) However, for rectangular grid graphs, DirectParition substantially outperforms GerryChain UST for most longer graph aspect ratios and outperforms MST on all but the longest aspect ratios (see Figure~\ref{rectGridGraphs}). The rectangular grid graphs may be especially relevant for ReCom where merged districts have irregular shapes. It is not surprising that our method excels against ordinary Wilson's algorithm in the elongated regime, as the expected lengths of our random walks are shorter, whereas those of Wilson's algorithm are longer.

As Python appeared to be constraining performance, we implemented a Cython version of the algorithm as well as a Cython implementation of a Wilson's bipartition to compare it to. Under Cython's reduced overhead, our approach outperformed the Cython Wilson's implementation across scales, even in the square regime (see Figure~\ref{cythonComparison}).

\ipncm{.55}{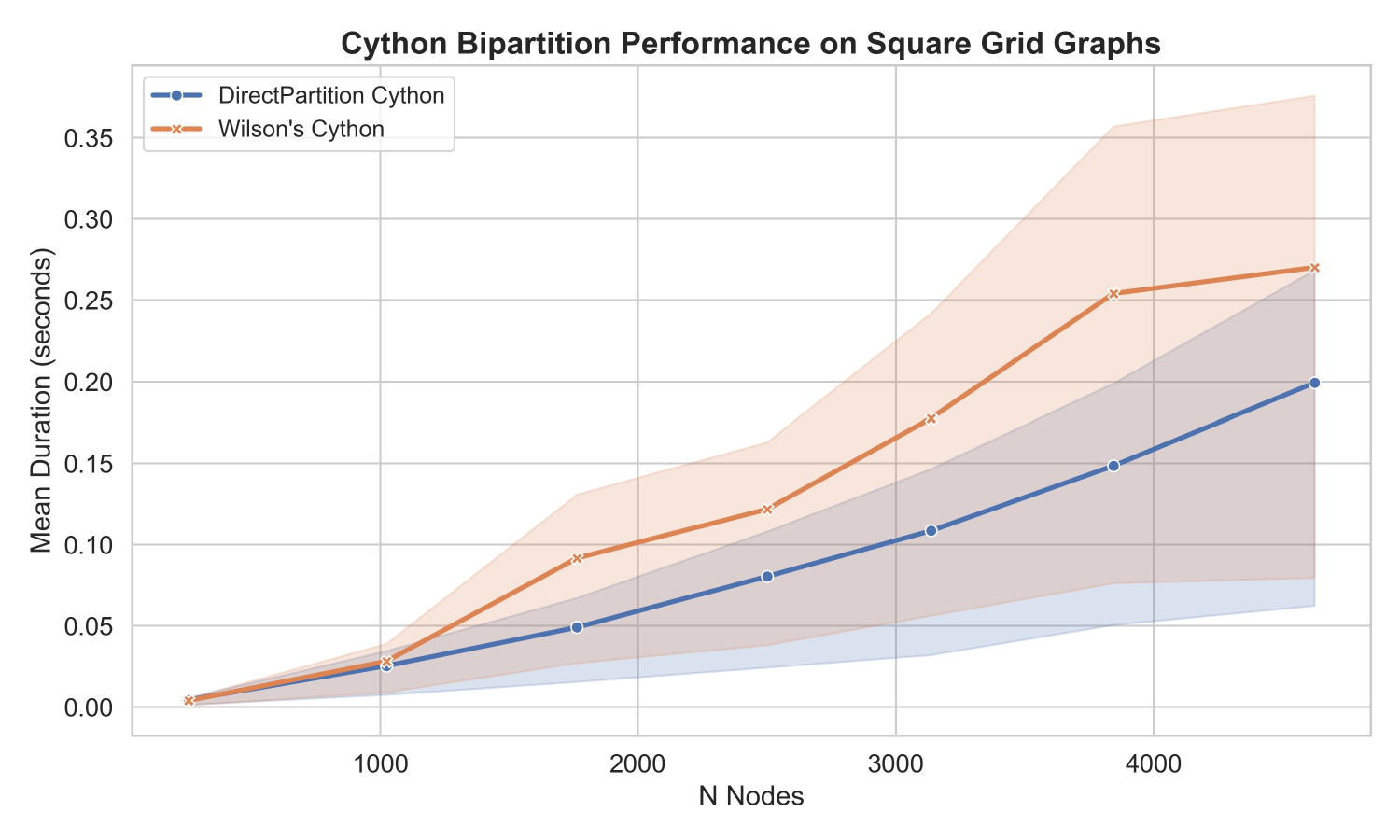}{\label{cythonComparison}Mean $2$-split time across grid-graph scales over $1000$ trials with $25^{th}$ and $75^{th}$ percentile bounds. (Benchmarked on Apple M1 Max.)}

\subsection{A Note on Simple Connectivity} \label{subNotsc}

Our definition of a grid-like graph requires that it be simply connected. This is because our algorithm runs in a planar graph's planar dual, where any hole would be contracted to a single vertex that potentially has very high degree. In this case simple random walks on the dual graph no longer behave well, with the single high-degree vertex potentially leading the random walk to quickly jump to different parts of the graph in an uncontrolled way. Our analysis relies critically on dual random walks being well-behaved (this is part of our definition of grid-like lattices), so holes are not amenable to this kind of analysis. 

Non-simply connected regions could occur with ReCom and RevReCom, when the union of two districts encircles additional district(s).  To avoid this, one can simply reject such steps: Never recombine two districts whose union isn't simply connected.\footnote{Alternatively, when the union of two districts is not simply connected, one could instead use Wilson's algorithm on its dual to draw the entire spanning tree of this union. There is no speed-up over prior work in this case.}  The strongest proof known for irreducibility of ReCom under tight balance conditions, which considers three districts in a triangular region of the triangular lattice~\cite{irredj}, does not use any large-scale recombinations of districts whose union isn't simply connected.  In this result, any recombinations of such districts involve only flip moves (when one vertex is reassigned to a new district) or recombination of vertices within a simply connected subset of the union, with all other district assignments remaining unchanged. This suggests rejecting moves with non-simply connected unions is unlikely to cause problems or significantly limit the ability of ReCom or RevReCom to explore the state space.

For empirical validation, we ran both ReCom and RevReCom on a $10 \times 10$ grid for 100,000 and 1 million steps, respectively (as RevReCom is known to have much larger convergence times that ReCom\footnote{Though RevReCom has a notoriously high self-loop probability due to various rejection filters, we verified that out of 1 million steps, an average of about 60,000 of them in each trial resulted in a new partition} \cite{RevReCom}). We considered both 5 districts and 10 districts, with exact balance required between all districts.  We counted both in how many plans there existed two districts whose union was not simply connected, and how frequently such districts were recombined.  The average across 10 trials are shown in Table~\ref{tab:exp}.

\begin{table}

\begin{centering}

    \begin{tabular}{|c|c|c|c|c|} 
    \hline
    &&& Average number of   & Recombinations of \\
    Method & Districts & Steps &plans with non-simply& non-simply \\
    &&& connected pair& connected pair\\\hline
    ReCom & 5 & 100,000 & 6.3 & 0  \\\hline
    ReCom & 10 & 100,000 & 10.6 & 0 \\\hline
    RevReCom & 5 & 1,000,000 & 74.2 & 0  \\\hline
    RevReCom & 10 & 1,000,000 & 78 & 0 \\\hline
    \end{tabular}
        \vspace{.2cm}
        
\end{centering}

    \caption{The average of 10 trials in a 10 $\times$ 10 grid graph.  Districts whose union is not simply connected are exceedingly rare, and it is even rarer still to recombine such a pair to produce new districts. The latter was never done in ReCom or RevReCom across all trials. Because RevReCom has much higher self-loop probabilities than ReCom, it was run for more steps to ensure ample exploration of the state space.}
    \label{tab:exp}
\end{table}

We see that have a pair of districts whose union is not simply connected is exceedingly rare, and recombining such a pair is rare still. Perhaps counterintuitively, when districts were larger, there were fewer instances of district unions not being simply connected. More importantly, across all 2 million steps of ReCom and 20 millions steps of RevReCom in our trials, two non-simply connected districts were never successfully recombined to produce new districts.  In particular, an implementation of one of these chains that used our algorithm when splitting a simply connected region and that used the slower, straightforward approach based on uniform random spanning trees otherwise would never have resorted to generating a uniformly random spanning tree in these example runs.
 
Finally, we note that checking if a grid-like graph $H$ contained in a grid-like lattice $\Lambda$ is simply-connected can be done in time $O(|V(H)|)$ by checking the structure of the faces in $H$ that form the boundary of $H$. This is easily done by looking at the duals $H^*$ and $\Lambda^*$. Thus such checks do not increase the asymptotic runtimes under consideration.

\section{Future Directions}\label{secConclusion}

Our result shows a theoretical separation between the complexities of generating a random spanning tree and generating the random partition it induces. It would of course be of significant interest to prove nontrivial lower bounds on the time required to sample uniformly random spanning trees; presently the comparison we can draw to our method is just to the best-known algorithms for sampling spanning trees.

Removing the requirement that the region be simply connected presents interesting theoretical challenges.  A natural approach is to induct on the number of holes, and use random walks in the initial phases simply to `cut out' the holes of the region, rather than make progress on the balance of the split.  The main challenge with this approach is understanding the probability that such walks will fail to escape a strangely shaped hole in a short amount of time---in particular, the high degree face of the hole in the dual graph creates an important local exception to the grid-like environment in which our random walks are taking place.

Finally, a fundamental question about the recombination Markov Chains is whether the total computational time that must be spent between successful recombination steps is polynomial in the number of vertices, for grid-like graphs. In this paper we have shown that the projection of a random forest to the space of balanced 2-partitions can be sampled in linear time for arbitrary simply connected subsets of grid-like graphs, but we have not addressed the probability of rejection due to the tree not beeing splittable into two balanced subtrees. While prior work~\cite{polytimeforests} shows this success probability is at least $1/\poly(n)$ in grid-like graphs, this result only applies when the region being divided has a ``nice'' boundaries (e.g., the analogy is to grids themselves, not to subsets of grids). Forthcoming work by Micah Gold~\cite{AimPL} shows that, even on square grids, there exist states from which all ReCom attempts have an exponentially small success probability. However, typical cases (e.g., states sampled from the stationary distribution) seem to take polynomial time, though this has still not been proven. Thus the average-case complexity of the ReCom transition function is still a wide-open question.

\bibliographystyle{plain}
\bibliography{bibliography}

\newpage\appendix
\section{Appendix: Deferred proofs}\label{appDeferredProofs}

\subsection{Proof of Proposition \ref{prop:zisagrid}}\label{appz2isagrid}

\propzisagrid*
\begin{proof}
    Properties \ref{itmStrongLatticeSequenceDegree}, \ref{itmStrongLatticeSequenceClose}, and \ref{itmStrongLatticeSequenceDense} are immediate.  Property \ref{itmStrongcircleproperty}, holds for any $r$ with $s = 8$ and $\rho = \frac18$.  Indeed: we divide the circle around any point $v$ into 8 equal-sized arcs by vertical, horizontal, and 45-degree diagonal lines. By the symmetry of the grid, a random walk from $v$ is equally likely to exit the graph in any arc, which means that, with probability at least $\rho = \frac18$ it leaves through any specific arc.

    For \ref{itmStrongLatticeSequenceEscape}, we will prove the stronger statement that for any set $S\subseteq \Lambda^*$, any $v\in S$, and any $p>0$ the probability that a random walk from $v$ is in the set $S$ at some fixed time $t_0\geq L|S|/p$ is at most~$p$. 

    We will use the following upper bound on the probability $p_t(x,y)$ that a random walk from the origin of the grid is at vertex $(x,y)$ on the $t$th step of the walk:
    \begin{equation}\label{xybound}
    p_t(x,y)\leq \frac{D}{t}e^{-(x^2+y^2)/(Ct)}
    \end{equation}
    (for fixed constants $C,D$).
    Note that a generalization of this statement for Cayley graphs was proved by Hebisch and Saloff-Coste \cite{gaussiangroupbound}.

    Now if $X_t$ is the position of the random walk at step $t$, we have 
    \[
    \Pr(X_t\in S)=\sum_{(x,y)\in S}p_t(x,y)\leq \sum_{(x,y)\in S}\frac{D}{t}e^{-(x^2+y^2)/(Ct)}.
    \]
    Now for $n=|S|,$ choose $r_n\approx \sqrt{n}/{\sqrt{\pi}}$ so that there are at least $n$ points within distance $r_n$ of the origin.  Observe that we have
    \begin{equation}\label{eqsumbound}
    \sum_{(x,y)\in S}\frac{D}{t}e^{-(x^2+y^2)/(Ct)}\leq \sum_{x^2+y^2\leq r_n^2}\frac{D}{t}e^{-(x^2+y^2)/(Ct)},
    \end{equation}
    since if $(x_0,y_0)$ is in $S$ but $x_0^2+y_0^2>r_n^2$, then some point $(x,y)$ with $x^2+y^2\leq r_n^2$ is \emph{not} in $S$, and replacing $(x_0,y_0)$ with $(x,y)$ increases the sum.

    We can bound the righthand side of \eqref{eqsumbound} by the corresponding integral:
    \begin{multline}
    \sum_{x^2+y^2\leq r_n^2}\frac{D}{t}e^{-(x^2+y^2)/(Ct)}\leq
    \frac{De^{1/(2C)}}{t}\iint_{x^2+y^2\leq r_n^2} e^{-(x^2+y^2)/(2Ct)}dxdy\\=
    \frac{2\pi De^{1/(2C)}}{t}\int_{r=0}^{r_n}r e^{-r^2/(2Ct)} dr=2\pi CDe^{1/(2C)}(1-e^{r_n^2/(2Ct)})\leq 2\pi CDe^{1/(2C)}(1-e^{-2n/\pi (2Ct)})\\\leq 2De^{1/(2C)}n/t,
    \end{multline}
    where here we have used that (crudely) $r_n^2\leq 2n/\pi$, and in the last step we have used that $e^{-x}\geq 1-x.$

    Now for $L=2De^{1/(2C)}$ and $t=L|S|/p$, we have that the probability that $X_n\in S$ is at most $Lnp/Ln=p,$ as claimed.
    \end{proof}

\subsection{Proof of Lemma \ref{lem:curveswalk}}\label{appcurveswalk}
\lemcurveswalk*
\begin{proof}
    This follows from applying \cite[Lemma 4.4]{polytimeforests} to the sequence $\Lambda_1, \Lambda_2, \Lambda_3, \dots$, where each $\Lambda_n$ is obtained by scaling down the drawing of $\Lambda$ (and its dual) in the plane by a factor of $n$. One can verify that $(\{\Lambda_n\}, \rho)$ is a \emph{lattice sequence},
    so the lemma applies, stating that, for any curve $\gamma'$ and any $\varepsilon > 0$, for all $n > \max\{\frac{100}{\varepsilon}, \frac{R}{\varepsilon}\}$, with probability at least
    $$\rho^{100\abs{\gamma'}/\varepsilon + 301},$$
    a random walk in $\Lambda^*_n$ from any $v_0$ such that $d(v_0, \gamma'(0)) \leq \varepsilon/2$ will encircle $\gamma'(1)$ without ever getting closer than $\varepsilon/5$ to this point and never reach a vertex at distance $> \varepsilon$ from $\gamma'$. Applying this to the curve $\gamma' := \frac{\gamma}{n}$ (which is $\gamma$ scaled scaled down by a factor of $n$), $\varepsilon := 1$, and $n := r$ (which is valid for $r \geq \max\{100, 10R\}$), we have that, with probability at least
    $$\rho^{100 \abs{\gamma'} + 301} = \rho^{100 \frac{\abs{\gamma}}{n} + 301} = \rho^{100 \frac{\abs{\gamma}}{r} + 301},$$
    a random walk from any $v_0 \in \Lambda^*_n$ such that $d(v_0, \gamma'(0)) \leq 1/2$ will encircle $\gamma'(1)$ without ever getting closer than $1/5$ to this point and never reach a vertex at distance $> 1$ from $\gamma'$. Scaling up by $r = n$, we have that, with probability at least $\rho^{100\abs{\gamma}/{r} + 301}$, a random walk from any $v_0 \in \Lambda^*$ such that $d(v_0, \gamma(0)) \leq r/2$ will encircle $\gamma(1)$ without ever getting closer than $r/5$ to this point and never reach a vertex at distance $> r$ from $\gamma'$. 
\end{proof}

\subsection{Proof of Lemma \ref{lemTreeCenter}}\label{appTreeCenter}

\lemTreeCenter*

\begin{proof}
First, we prove that every vertex-weighted tree has at least one vertex center by giving an algorithm to find one.  Let $T$ be a tree and let $v$ be any vertex of $T$. For convenience, let $W$ denote the total weight of all the vertices in $T$. Look at the sizes of the subtrees of $v$ (that is, the components of $T \setminus v$).   If all subtrees of $T \setminus v$ have weight at most $W/2$, then $v$ is a center and we are done.  Otherwise, at least one subtree $T'$ must have cumulative  weight greater than $W/2$.  As it is not possible for two subtrees to have total weight greater than $W/2$, then $T'$ is the unique subtree of $v$ with weight greater than $W/2$.  Let $x$ be $v's$ neighbor in $T'$.   Propose $x$ as the new possible vertex center and repeat.  Note that one component of $T \setminus x$ is $v$ together with its subtrees that are not $T'$.  As $T'$ has weight greater than $W/2$, this new component has weight less than $W/2$. This means the algorithm will not return to $v$ in a future step, meaning as the tree has a finite number of vertices this process will eventually terminate.  The vertex at which it terminates must be a vertex center.  

This algorithm takes linear time:  First, root $T$ at $v$ and use a postorder traversal to recursively compute the size of each vertex's subtree. In each iteration, it then takes $O(1)$ time to determine whether a vertex is a center, and there are at most $O(n)$ iterations. Thus a vertex center always exists and can always be found in linear time. 

Note that if $T$ has an edge center, then both of its endpoints must be vertex centers.  Once a vertex center is found, one simply needs to check all its adjacent edges to see if they are edge centers. Thus an edge center, if one exists, can also be found in linear time. 

Finally, note that if $T$ has multiple vertex centers, they must be adjacent.  If there are two adjacent vertex centers, then the edge between them must also be an edge center.  We conclude that if there is no edge center, the vertex center is unique. 
\end{proof}

\subsection{Proof of Proposition \ref{proSpeedup}}\label{appSpeedup}

\proSpeedup*
\begin{proof}
Suppose there is a 0-splittable spanning tree $S$ extending $R(T^*)$. Let $(S_1, S_2)$ be the two subtrees of $S$ of equal size obtained when deleting some edge $e$. Note because $A$ is a vertex center, $e$ must either be a bridge with one endpoint in $A$ or it must have both endpoints in $A$; if this was not the case, $A$ could not possibly be a center of $R(T^*)$. 

Let $e'$ be a bridge $e'$ leaving $A$; consider all the vertices on the opposite side of $e'$ from $A$, that is, the subtree rooted at the endpoint of $e'$ that is not in $A$. It is not possible for some of these vertices to be in $S_1$ and some of these vertices to be in $S_2$, as this set is connected and $S$'s balance edge $e$ cannot have both its endpoints in this set. Therefore all the vertices in this subtree are in $S_1$ or are in $S_2$.  It follows that we can partition the neighbors of $A$ into two sets $A_1$ and $A_2$ such that the subtrees of all vertices in $A_1$ are in $S_1$ and the subtrees of all vertices in $A_2$ are in $S_2$.  As we know $S_1$ and $S_2$ each contain exactly half of the total vertices in $H$, it follows that the total weight of the vertices in the subtrees of each $A_i$ is at most half the total weight. 

It only remains to show that $A_1$ and $A_2$ are contiguous in the cyclic order of the embedding. This follows immediately from planarity and the fact that all connections between vertices of $A_1$ and vertices of $A_2$ must occur within $R$: if these subtrees were not cyclically contiguous according to the embedding, it would be impossible for both $S_1$ and $S_2$ to be connected. \end{proof}

\end{document}